\documentclass[5p]{elsarticle}
\usepackage{amssymb,amsmath}
\usepackage{lineno}
\usepackage{natbib}
\usepackage{color}
\usepackage{url}
\usepackage{multirow}
\usepackage{graphicx}
\usepackage{subfigure}
\usepackage{bm}
\journal{Nuclear Instruments and Methods in Physics Research A}
\usepackage{enumitem}
\usepackage{siunitx}
\usepackage[]{algorithm2e}

\usepackage{appendix}

\usepackage{float}

\begin{document}
\begin{frontmatter}

\title{Polynomial Chaos Expansion method as a tool to evaluate and quantify field homogeneities of a novel waveguide RF Wien Filter}
\author{J. Slim$^1$, F. Rathmann$^2$, A. Nass$^2$,  H. Soltner$^4$, R. Gebel$^2$, J. Pretz$^{3,5}$ and D. Heberling$^{1,5}$}
\address{
$^1$ Institute of High-Frequency Technology, Rheinisch-Westf{\"a}lische Technische Hochschule Aachen, 52074 Aachen, Germany\\
$^2$ Institute of Nuclear Physics (IKP), Forschungszentrum J{\"u}lich GmbH, 52428 J{\"u}lich, Germany\\
$^3$ III. Physikalisches Institut B,  Rheinisch-Westf{\"a}lische Technische Hochschule Aachen, 52056 Aachen, Germany\\
$^4$ Central Institute of Engineering and Analytics (ZEA), Forschungszentrum J{\"u}lich GmbH, 52428 J{\"u}lich, Germany\\
$^5$ JARA-FAME (Forces and Matter Experiments), Forschungszentrum J{\"u}lich and RWTH Aachen University, 52056 Aachen, Germany\\}

\begin{abstract}

For the measurement of the electric dipole moment of protons and deuterons, a novel waveguide RF Wien filter has been designed and will soon be integrated at the COoler SYnchrotron at J\"ulich. The device operates at the harmonic frequencies of the spin motion. It is based on a waveguide structure that is capable of fulfilling the Wien filter condition ($\vec{E} \perp \vec{B}$) \textit{by design}. The full-wave calculations demonstrated that the waveguide RF Wien filter is able to generate high-quality RF electric and magnetic fields. In reality, mechanical tolerances and misalignments decrease the simulated field quality, and it is therefore important to consider them in the simulations. In particular, for the electric dipole moment measurement, it is important to quantify the field errors systematically. Since Monte-Carlo simulations are computationally very expensive, we discuss here an efficient surrogate modeling scheme based on the Polynomial Chaos Expansion method to compute the field quality in the presence of  tolerances and misalignments and subsequently to perform the sensitivity analysis at zero additional computational cost.  
 
\end{abstract}

\end{frontmatter}

\section{Introduction and motivation} \label{sec1}
The results of simulated electromagnetic models and the real fabricated systems may differ. Not every single detail can be included in the electromagnetic model due to the finite computational capacity. Assumptions are made and some aspects of the real-world model are ignored. The electromagnetic models  themselves are error-prone,  depending for instance on the number of mesh cells and discretization uncertainties.

Fabrication processes, with their inherent mechanical uncertainties, have an impact on the actual electromagnetic response. Waveguides are one example of such systems; the electromagnetic modes are influenced as a result of the mechanical uncertainties. The RF Wien filter recently designed is based on a parallel-plates waveguide\,\cite{Slim2016116}. This waveguide is subject to stochastic mechanical variations. Unfortunately, analytic solutions for the fields resulting from a complex structure such as the RF Wien filter do not exist. Approximation methods such as perturbation theory for instance\,\cite{simmonds2013first} cannot be applied to tackle tolerance problems, therefore numerical solutions are applied to quantify the uncertainties.

The classical Monte-Carlo (MC) approach requires the calculation of a large number of responses corresponding to a set of uncertain parameters. 
Basically, Monte-Carlo methods converge  slowly in the sense that they require a substantial number of simulations (typically of the order of $10^4$ to $10^5$) to provide a reliable estimate of the system performance. We rely on a commercially available software to simulate the RF Wien filter\footnote{CST Microwave Studio - Computer Simulation Technology AG, Darmstadt, Germany, \url{http://www.cst.com}}. In the case where the model evaluation requires full-wave simulations of a complex electromagnetic structure such as the RF Wien filter, the Monte-Carlo approach cannot be used because of the long time required for each simulation.

The proposed method in this study is primarily based on the so-called Polynomial Chaos Expansion (PCE). The aim is to build a surrogate mathematical model that requires only a small number of model evaluations to reconstruct the response of the RF Wien filter in terms of the field homogeneity. Surrogate modeling is one possible solution that is feasible and which provides a reliable method to quantify uncertainties and provides a sensitivity analysis at zero additional cost. The PCE method has been used in many engineering applications, such as modeling uncertainties in electric motors\,\cite{7093523}, and also in accelerator science\,\cite{Adelmann:2015wva}.

This paper is organized as follows: Section\,\ref{sec2} describes the motivation and the requirements for the proposed modeling scheme. Section\,\ref{sec3} describes the considered electromagnetic field uncertainties. Section\,\ref{sec4} briefly introduces the PCE theory and describes the mathematical formulation necessary for the purpose of this paper. Section\,\ref{sec5} shows the application of the PCE theory and how it can be used to quantify the quality of the electric and magnetic fields. To conclude the analysis, Section\,\ref{sec6} reports about the global PCE-based sensitivity analysis, and in Section\,\ref{sec7} results are summarized. Four appendices can be found at the end of the paper. \ref{app:1} explains the construction of the univariate and multivariate Hermite orthogonal polynomials (basis), and\,\ref{app:2} clarifies and validates the theory with a simple example. \ref{app:3} discusses the employed PCE truncation scheme. Finally, \,\ref{app:4} explains the least-angle regression method used in this paper.

\section{Modeling mechanical uncertainties of the RF Wien filter} \label{sec2}
\begin{figure}[hbt]
\centering
\includegraphics[width=8.5cm]{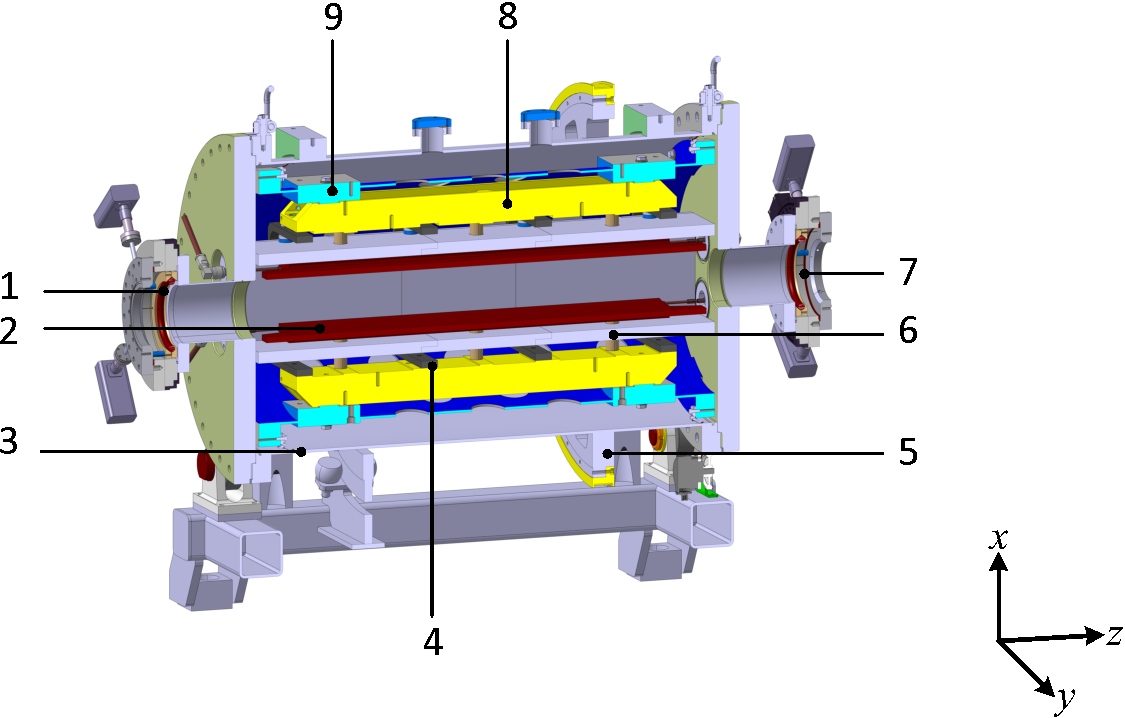}
\caption{\label{fig:structure} CAD design model of the RF Wien filter, showing the parallel-plates waveguide and the support structure (Figure taken from\,\cite{Slim2016116}). The coordinate system used for the design calculations is indicated. The stored beam moves along the $z$-axis of the RF Wien filter. 1: beam position monitor (BPM); 2: copper electrodes; 3: vacuum vessel; 4: clamps to hold the ferrite structure; 5: belt drive for $\ang{90}$ rotation; 6: ferrite structure; 7: CF160 rotatable flange; 8: support structure of the electrodes; 9: inner support tube.}
\end{figure}

The RF Wien filter is based on a novel concept of a parallel-plates waveguide as shown in Fig.\,\ref{fig:structure}. 
The system consists of a transmission line composed of two conductors that supports the transverse electromagnetic (TEM) mode in the frequency range required by the electric dipole moment experiment (see\,\cite{Slim2016116}). The RF Wien filter requires an orthogonal electromagnetic field without components in the beam direction which is fulfilled by the TEM mode. The field quotient $Z_q= -E_x/H_y$ controls the Wien filter condition (see\,\cite{Slim2016116}), \textit{e.g.}, for deuterons at $\SI{970}{MeV \per c}$, $Z_q$ must be $\approx \SI{173}{\ohm}$ to provide zero Lorentz force. To set $Z_q$ to any particular value, the wave mismatch theory\,\cite{slimpstp2015} has been used. This mismatch forces part of the electromagnetic field to be reflected back into the structure, thereby forcing a forward and backward propagation of fields. The fields sum up vectorially to produce the necessary field quotient. Together with the optimally shaped electrodes, a minimal integral Lorentz force can be ensured. A ferrite structure surrounds the electrodes to homogenize the magnetic field and to increase the magnetic field by 25\% compared to a ferriteless solution.
\begin{figure}[t]
\centering
\includegraphics[width=8cm]{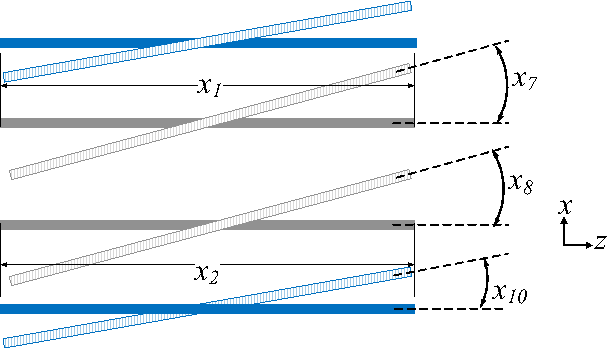}
\caption{Stochastic variation of the parameters of the waveguide RF Wien filter in the $xz$-plane. The solid grey lines represent the electrodes and the  solid blue lines the ferrites. The shaded counterparts indicate the possible rotational displacements and misalignments. $x_1$ and $x_2$ model the random lengths while $x_7$ and $x_8$ model the random angular rotations of the electrodes in the $xz$-plane. $x_{10}$ models the possible rotation of the ferrites.}
\label{fig:pce1}
\end{figure}
\begin{figure}[t]
\centering
\includegraphics[width=7cm]{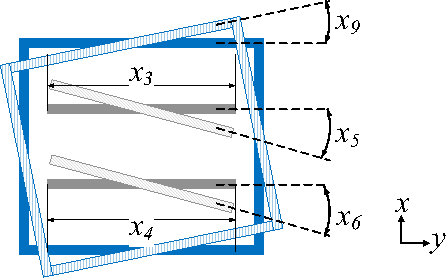}
\caption{Stochastic variation of the parameters of the waveguide RF Wien filter in the $xy$-plane. The solid grey lines represent the electrodes and the solid blue lines the ferrite structure.  The shaded counterparts indicate the possible rotational displacements and misalignments. $x_3$ and $x_4$ model the random widths of the electrodes while $x_5$ and $x_6$ model the random angles rotations of the electrodes in the $xz$-plane. $x_9$ models the possible rotation of the ferrites.}\label{fig:pce2}
\end{figure}

As shown in Fig.\,\ref{fig:pce1}, $x_1$ and $x_2$ represent the lengths of the upper and lower electrode, respectively. $x_1$ and $x_2$ are considered independent random variables. According to the information provided by the manufacturer\,\cite{zea}, the electrodes can be produced with a fabrication tolerance of $\pm \SI{0.1}{mm}$, such that $x_{1,2} = \SI{808.8}{}\pm \SI{0.1}{mm}$. Statistically, $x_1$ and $x_2$ are modeled as Gaussian-distri\-buted, independent random variables with a mean $\mu = \SI{808.8}{mm}$ and a standard deviation $\sigma = \SI{0.1}{mm}$, abbreviated as $G\left( \mu, \sigma \right)$. 

\begin{table*}[hbt]
\centering
\caption{Statistical distributions of the random variables representing the stochastic mechanical variations of the waveguide RF Wien filter. $G \left(\mu, \sigma \right)$ indicates a Gaussian distribution with a mean value of $\mu$ and a standard deviation of $\sigma$, in units of \si{mm} or \si{mrad}.}
\renewcommand{\arraystretch}{1.2}
\begin{tabular}{llll}\hline \hline
Variable & Description & Distribution & Unit\\\hline
$x_1$	 & length of the upper electrode			& $G \left(808.8,0.1\right)$ 	&	\si{mm}\\
$x_2$	 & length of the lower electrode			& $G \left(808.8,0.1\right)$ 	&	\si{mm}\\
$x_3$	 & width of the upper electrode				& $G \left(182,0.1\right)$   	&	\si{mm}\\
$x_4$	 & width of the lower electrode				& $G \left(182,0.1\right)$   	&	\si{mm}\\
$x_5$	 & rotation of the upper electrode in the ($xy$-plane) 	& $G \left(0,1\right)$		&	\si{mrad}\\
$x_6$	 & rotation of the lower electrode in the ($xy$-plane) 	& $G \left(0,1\right)$		&	\si{mrad}\\
$x_7$	 & rotation of the upper electrode in the ($xz$-plane) 	& $G \left(0,1\right)$		&	\si{mrad}\\
$x_8$	 & rotation of the lower electrode in the ($xz$-plane) 	& $G \left(0,1\right)$		&	\si{mrad}\\
$x_9$	 & rotation of the ferrite structure in the($xy$-plane)     	& $G \left(0,1\right)$		&	\si{mrad}\\
$x_{10}$ & rotation of the ferrite structure in the($xz$-plane)     	& $G \left(0,1\right)$		&	\si{mrad}\\
\hline \hline
\end{tabular}
\label{tab:var}
\end{table*}

In reality, deviations from the ideal parallelism of the electrodes may occur which needs to be included in the field simulations. This is taken into account by allowing the two plates to rotate in the $xy$ and $xz$ plane (see Fig.\,\ref{fig:pce1} and \,\ref{fig:pce2} for coordinates). A rotation in the $yz$ plane is less probable and is therefore not considered here\footnote{Each electrode is connected to the support structure via eight metallic screws. For a rotation to occur, all screws would have to be simultaneously misaligned which is unlikely.}. 
$x_7$ and $x_8$ represent the tilt angles of the possible rotations of the electrodes in the $xz$ plane (see Fig.\,\ref{fig:pce1}). The electrodes can also move independently with a rotational interval up to $\pm \SI{1}{mrad}$\,\cite{zea}. 

As shown in Fig.\,\ref{fig:pce2}, the variation of the width of the electrodes ($x_3$ and $x_4$) and their rotation in the $xy$-plane ($x_5$ and $x_6$) are considered as well in the calculations. $x_9$ and $x_{10}$ correspond to undesired rotations of the ferrite structure in the $xy$ and the $xz$ plane, respectively. All random variables may vary simultaneously. The statistical characteristics of the random variables considered here are listed in Table\,\ref{tab:var}. 

\section{Electromagnetic field uncertainties} \label{sec3}

The desired electric field should point into the $x$-di\-rec\-tion, thus ideally produce $\left(E_x, 0, 0\right)$,  while the magnetic field should point into the $y$-direction,  $\left(0,H_y,0\right)$. The unwanted field components are denoted by $\vec{E}_\perp$ and $\vec{H}_\perp$ and were defined in\,\cite{Slim2016116} as
\begin{eqnarray}
\vec{E}_\perp &=& \left(\begin{array}{ccc}
  0\\
  E_y \\
  E_z\\
 \end{array}\right) \label{eq:e_perp} \,, \, \text{and}\\
\vec{H}_\perp &=& \left(\begin{array}{ccc}
  H_x\\
  0 \\
  H_z\\
 \end{array}\right) \label{eq:H_perp}. \,
\end{eqnarray}

\begin{figure*}[hbt]
\centering
\subfigure[1000 full-wave simulations showing the magnitude of the unwanted electric field $|\vec E_\perp|$ along the beam axis. ]
{\includegraphics[width=0.45\textwidth]{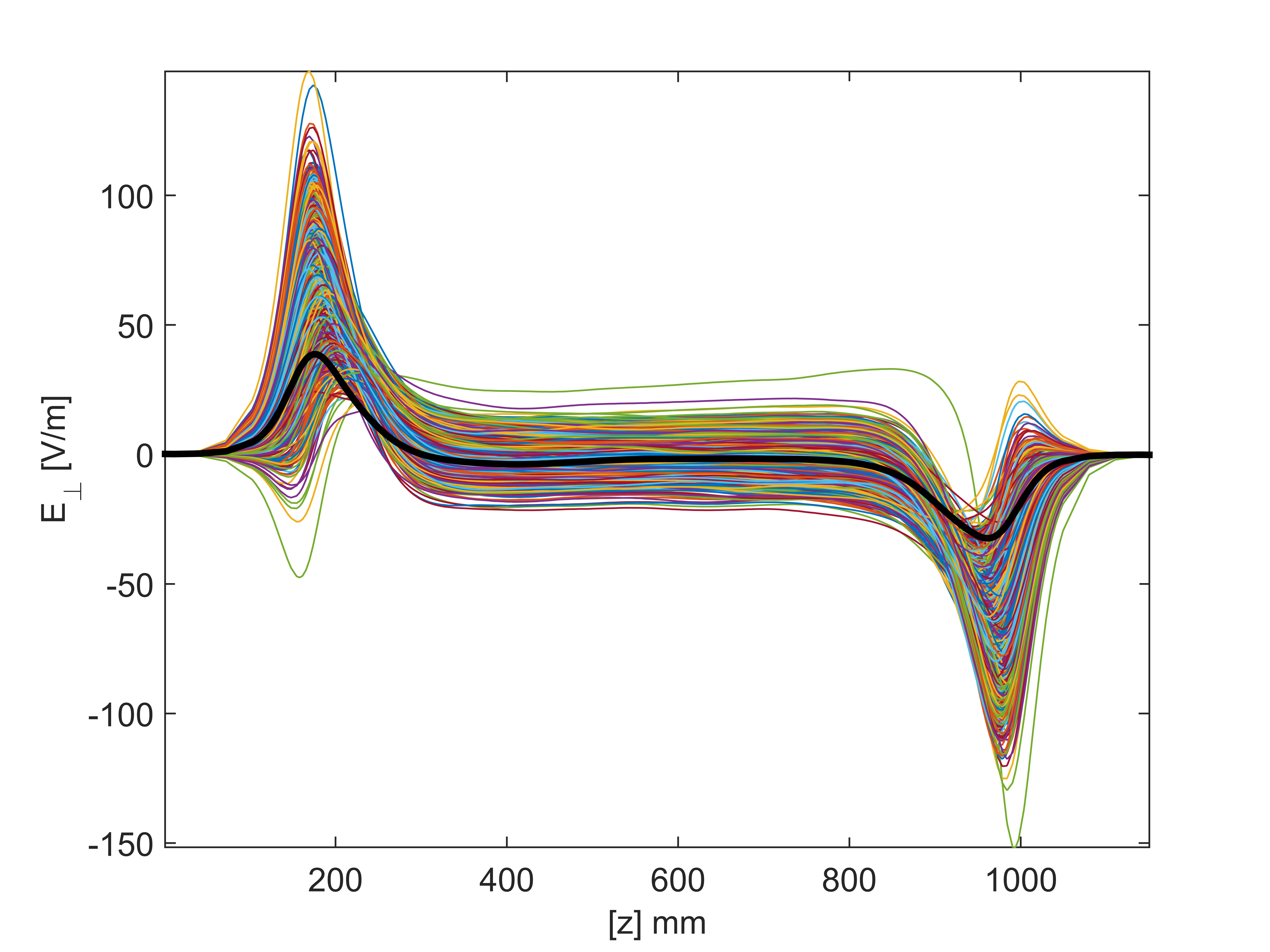}}
\hspace{0.5cm}
\subfigure[1000 full-wave simulations showing the magnitude of the unwanted magnetic field $|\vec H_\perp|$ along the beam axis.]  
{\includegraphics[width=0.45\textwidth]{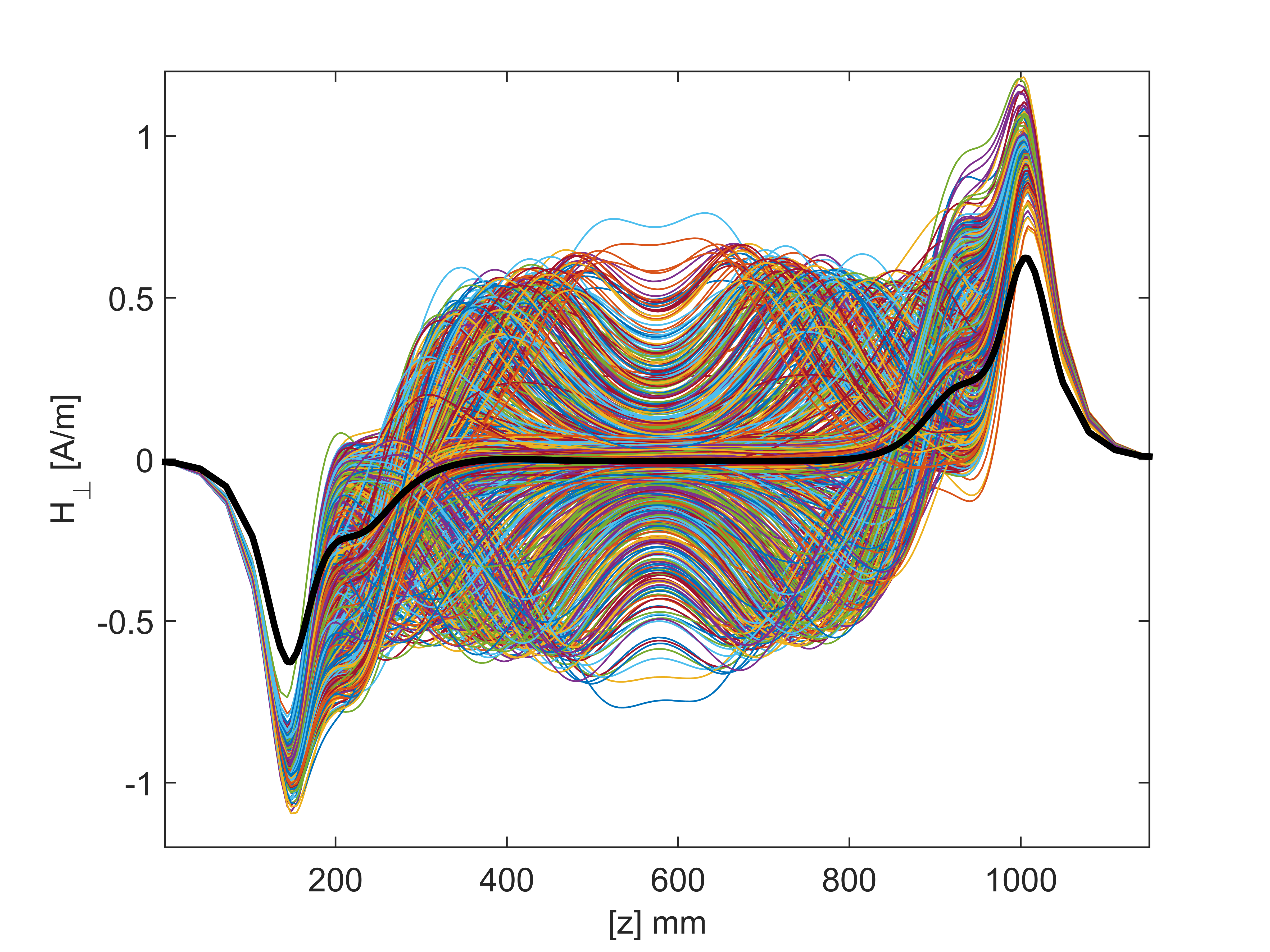}}
\caption{\label{fig:eh_perp} Low-order Monte-Carlo simulations showing the evaluation of $|\vec E_\perp|$ and $|\vec H_\perp|$ along the beam axis for a $\SI{10}{kW}$ input power. In the worst case, $| \vec E_\perp|$ does not exceed $\SI{150}{V/m}$, while $|\vec H_\perp|$ does not exceed $\SI{1}{A/m}$. The solid black line represents the ideal case, \textit{i.e.}, with perfect alignment of all elements of the RF Wien filter and zero tolerances.}
\end{figure*}

The results for 1000 full-wave simulations are collected in Fig.\,\ref{fig:eh_perp}, in terms of the magnitude of $\vec{E}_\perp$ and $\vec{H}_\perp$. These simulations used the Gaussian distribution of the $x_i$ ($i=1,\ldots,10$), as given in Table\,\ref{tab:var}, and the stochastic variations of the unwanted fields $E_\perp$ and $H_\perp$ are evaluated along the beam axis. Mechanical tolerances lead to deviations of the wave vector $\vec{k}$ from its ideal direction along the beam axis. Such deviations generate unwanted fields. For instance $E_\perp$, shown in Fig.\,\ref{fig:eh_perp}(a), can reach $\SI{30}{V/m}$ inside the RF Wien filter, while at the edges, a value $\SI{150}{V/m}$ is possible. The fields at the edges are approximately 5 times larger than the inner ones. 

The magnetic fields $\vec{H}_\perp$, shown in Fig.\,\ref{fig:eh_perp}(b), exhibit a different behavior. The unwanted fields at the edges have roughly twice the magnitude compared to the inner ones.  Inner deviations reach up $\SI{0.7}{A/m}$ in contrast to $\SI{1}{A/m}$ at the entrance and exit of the RF Wien filter. Uncertainties deform the cross-section of the waveguide at the entry and exit points of the Wien filter in an asymmetric manner, resulting in non-equidistant and non-parallel electrodes. This explains the non-uniformity and asymmetries of the field variations. 

The field homogeneities, specified via $f_{E_\perp}^{\text{int}}$ and $f_{H_\perp}^{\text{int}}$ are calculated by taking into account the total fields (see\,\cite{Slim2016116}), which results in    
\begin{equation}
\begin{split}
f_{E_\perp}^{\text{int}} & = \frac{\int_{-\ell/2}^{\ell/2}|\vec{E}_\perp| d \ell}{\int_{-\ell/2}^{\ell/2}|\vec{E}| d \ell} = \mathcal{Y}_E\,, \text{ and}\\
f_{H_\perp}^{\text{int}} & = \frac{\int_{-\ell/2}^{\ell/2}|\vec{H}_\perp| d \ell}{\int_{-\ell/2}^{\ell/2}|\vec{H}| d \ell}= \mathcal{Y}_H \,.
\end{split}
\label{eq:f-eh_perp}
\end{equation}
The effective length of the RF Wien filter amounts to $\SI{1550}{mm}$. In order to speed up the calculations, the integrals in Eq.\,(\ref{eq:f-eh_perp}) were evaluated for $\ell=\SI{1152}{mm}$ along the longitudinal axis of the RF Wien filter. The field contributions outside the considered length are very small (of the order of $\num{e4}$) and are not taken into account. In the following, the electric and magnetic fields homogeneities are denoted by $\mathcal{Y}_E$ and $\mathcal{Y}_H$. These are the quantities that this paper tries to estimate within a reasonable number of full-wave simulations.

\subsection{Note on the full-wave simulations} 
During the design phase of the RF Wien filter, around $\num{63e6}$ mesh cells have been used in the simulation software. Each simulation required roughly 12 hours on a GPU-based computing system. For a reliable MC simulation, $10^5$ simulations would be required, corresponding to years of computation time. Clearly this is not a feasible solution. 

For the uncertainty analysis, the complexity of the electromagnetic model of the waveguide RF Wien filter has been reduced, ignoring therefore many details of the mechanical model, \textit{e.g.}, support screws and their holding rings. The number of mesh cells has been reduced to $\num{38e6}$ mesh cells. Moreover, the material properties of some components have been simplified; the vacuum vessel and the inner support tube do not affect the field quality and they can be safely changed from non-magnetic stainless steel to PEC (perfect electric conductors).  Thereby, the number of mesh cells is further decreased to $\num{13e6}$ and the simulation time for one configuration is reduced to nearly $\SI{2.5}{\hour}$. 

$f_{E_\perp}^{\text{int}}$ and $f_{H_\perp}^{\text{int}}$ have been computed. Compared to the complex model, the reduced-order model exhibits small differences, but it should be noted that without such simplification, an uncertainty analysis would not be possible at all. Computation times of $\SI{2.5}{\hour}$ are also not very affordable ($\SI{2.5}{\hour}$ is the pure solving time, which does not include the preparation phase for the simulations, \textit{e.g.}, the meshing and the preparation of the material matrices). In total, it took more than $\SI{115}{days}$ to perform 1000 simulations, which were carried out discontinuously, depending on the availability of the GPU-cluster.   

\section{Polynomial chaos expansion (PCE)} \label{sec4}
\subsection{Introduction}

Polynomial chaos expansion (PCE) is a spectral method that can describe randomness (\textit{e.g.}, uncertainties) in stochastic dynamical systems in a Fourier-like series expansion. It was originally proposed by Norbert Wiener\,\cite{10.2307/2371268} and integrated into the finite-element method (FEM) by Ghanem and Spanos\,\cite{Ghanem:1991:SFE:103013}.

Formally, for the polynomial chaos expansion to be defined, a \textit{complete} probability space is required. Let $(\Omega, \mathcal{F}, \mathcal{P})$ be the \textit{complete} probability space, where $\Omega$ is the sample space, $\mathcal{F}$ is the $\sigma$-field and $\mathcal{P}$ is the probability measure on $\mathcal{F}$\,\cite{ash2000probability}. Wiener found that the Hermite polynomials are orthogonal with respect to Gaussian probability measures $\mathcal{P}$ in $\mathcal{L}^2$-normed $(\Omega, \mathcal{F}, \mathcal{P})$ spaces and can therefore be used as basis functions to span \textit{Gaussian spaces}. This allows \textit{any} \textbf{random} process with finite second-order moment (regardless of what it represents) to be fully described by a set of \textbf{deterministic} coefficients (see the Cameron-Martin theorem for more details in\,\cite{lifshits2012lectures}). In principle, the theory is not restricted to Gaussian spaces, in fact, other polynomials can span non-Gaussian spaces as described in\,\cite{askeyWiener}. 

\subsection{System representation and calculation of coefficients}

Fundamental for the PCE calculations is the so-called multi-index set. This is simply a multi-dimensional indexing scheme related to the tensor product of the Hermite polynomials. The tensor product is one of the methods that can be utilized to produce the multi-dimensional Hermite polynomials. The multi-index set is represented as $\bm{i}$ and is defined as 
\begin{equation}
\mathcal{I}_{m,p}=\Big\{\bm{i}=[i_1,\cdots,i_m]; \parallel\bm{i}\parallel_1\le p \Big\}
\label{eq:multi_i}
\end{equation}
$p$ is the order of the polynomial (also called the expansion order), $\parallel\parallel_1$ is the $\mathcal{L}^1$ norm,  explained in\,\ref{app:2}. $\bm{x}$ is the vector of all random parameters that represent the possible uncertainties in the RF Wien filter, \textit{i.e.}, $\bm{x} = \{x_1, x_2,...,x_{10}\}$, as collected in Table\,\ref{tab:var}. $\mathcal{Y}$ denotes the output of the system (also called the model response). $\mathcal{Y}$ describes in fact the electric and magnetic field inhomogeneities. 

For the PCE series to converge, it is necessary that $\bm{x}$ be standardized. Standardization is a form of iso-probabilistic transformation that transforms an arbitrary random variable into a normally distributed one with zero mean and unity standard deviation. Thus, the input parameters $x_i$ are transformed into the corresponding standardized random variables $\xi_i$. The stochastic spectral representation, expanding the model response $\mathcal{Y}$ with an $m$-dimensional truncated PCE to the order $p$ reads
\begin{equation}
\mathcal{Y}=\mathcal{M}\left( \bm{\xi}  \right) = \sum_{\bm{i}\in \mathcal{I}_{m,p}} \alpha_{\bm{i}} \Psi_{\bm{i}} \left( \bm{\xi} \right)\,.
\label{eq:main_pc}
\end{equation}
Here $\mathcal{M}$ is called the meta-model and it denotes the solver used. In the analysis conducted by Ghanem and Spanos\,\cite{Ghanem:1991:SFE:103013}, $\mathcal{M}$ was the finite-element method (FEM). But it can be in principle any solver such as the finite integration technique (FIT), as in the case of this work\footnote{The deterministic code used in this work is the full wave simulator commercial software package, CST Microwave Studio. The simulations were executed on a dual-Xeon E5 CPU with 4 Tesla C2075 GPU with 448 CUDA cores each.}. The $\alpha_{\bm{i}}$ denote the (\textit{deterministic}) expansion coefficients to be determined, $\Psi_{\bm{i}}$ are the basis functions (the multivariate chaos polynomials). The detailed computation of the homogeneous chaos basis functions is explained in\,\ref{app:1}.

To compute the expansion coefficients, intrusive and non-intrusive approaches can be used. Intrusive methods alter the underlying code by introducing another form of discretization. This \textit{stochastic} discretization converts the governing stochastic equation into a large system of linear equations. This technique is computationally fast, but altering code is often error-prone. In addition, the deterministic code may not be available, or may require many difficult analytical calculations. The other approach, the non-intrusive one, considers the (existing) deterministic code as a black-box. No changes of the code are required. It runs independently from the solver used in the deterministic code, which makes it an attractive option. The expansion coefficients are calculated having multiple calls to the deterministic code via \textit{regression} or \textit{projection}. 

\textit{Projection} requires the evaluation of expected values (integrals) and relies on the orthogonality of polynomials to compute the coefficients in the form of
\begin{equation}
\alpha_{\bm{i}} = \frac{\mathbf{E}\left\{\mathcal{Y}\Psi_{\bm{i}}\right\}}{\mathbf{E}\left\{\Psi_{\bm{i}}^2 \right\} }\,.
\label{eq:proj}
\end{equation}

\textit{Regression} on the other hand computes the coefficients with the least-square minimization method. According to\,\cite{OffermannDiss}, regression leads to more accurate results and it is used in this work. Regression methods require the solution of a large system of linear equations (LLSE) and matrix inversion, as will be shown later. 

Matrix inversion may fail. To avoid this case, a well-posed restriction is imposed on the LLSE, in the least-square sense. In terms of the PCE analysis, this restricts the lower bound on the costly lower number of full-wave deterministic simulations $N$. $N$ as a rule of thumb must be at least 1.5 times the number of polynomials basis $P$ calculated as the following permutations
\begin{align}
         P=\begin{pmatrix}
           m+p \\           
           p \\
          \end{pmatrix} = \frac{\left( m+p\right)!}{m!}\,.
          \label{eq:permu}
     \end{align}

Rewriting Eq.\,(\ref{eq:main_pc}) in matrix form, yields 
\begin{align}
\label{eq:main_pc_matrix}
\left[ \begin{array}{c} \mathcal{Y}^1 \\\\ \mathcal{Y}^2\\\\ \vdots \\\\ \mathcal{Y}^N \end{array} \right] =  
\left[
\begin{matrix}
\Psi_0 \left( \bm{\xi_1}\right)  & \dots & \Psi_0 \left( \bm{\xi_N}\right) \\\\
\Psi_1 \left( \bm{\xi_1}\right)  & \dots & \Psi_1 \left( \bm{\xi_N}\right) \\\\
\vdots  & \ddots &  \vdots \\\\
\Psi_{P-1} \left( \bm{\xi_1}\right)  & \dots & \Psi_{P-1} \left( \bm{\xi_N}\right) \\
\end{matrix}\right]^T \cdot
\left[ \begin{array}{c} \alpha_1 \\\\ \alpha_2\\\\ \vdots \\\\ \alpha_N \end{array} \right]\,.
\end{align}
Equation\,(\ref{eq:main_pc_matrix}) must be solved for the electric and magnetic field inhomogeneities individually.

By using the regression method, and assuming $N$ is large enough, the coefficients are calculated using the least-square estimation method (LSE), given by
\begin{equation}
\alpha_{\bm{i}} =\left(\bm{\Psi} \cdot \bm{\Psi}^T \right)^{-1}\cdot \bm{\Psi} \cdot \mathcal{Y}.
\label{eq:olse}
\end{equation}

The output of the full-wave simulations is expressed in terms of the electric and magnetic field homogeneities via $\mathcal{Y}_E$ and $\mathcal{Y}_H$, respectively. $\mathcal{Y}$ without any index refers to either of the two cases.  

\subsection{Sparse PCE}

With high-dimensional problems (the number of uncertain variables $m \ge 10$), such as the RF Wien filter, classical-full rank PCE is not feasible in the sense that the number of full wave simulations is intolerably large. In this case, PCE theory does not provide any advantages compared to MC methods. This was the motivation to look for a sparse version of PCE. Blatmann et al.\,\cite{blatman2010adaptive} proposed a systematic methodology to build an adaptive sparse PCE. Blatmann proposed to select \textbf{only} a subset of the basis functions that have the highest effects on the system response and to reject the other functions using a two-step procedure, in particular, a hyperbolic truncation scheme followed by a least angle regression (LAR) algorithm\,\cite{efron2004}. In his dissertation, Blatman showed that sparse PCE can still produce accurate meta-models but with much less complexity.

The PCE series shown in Eq.\,(\ref{eq:main_pc}) is truncated\footnote{Any expansion implemented on a computer is actually truncated.} up to a finite order $p$. The \textbf{hyperbolic truncation scheme}\,\cite{blatman2011adaptive}, implements an additional truncation scheme on the already $p$-truncated series by eliminating the basis functions with the highest order of interaction, as clarified in\,\ref{app:3}. In\,\cite{masonDoS}, an \textit{interaction} is formally defined as the "\textit{Existence of joint factor effects in which the effect of each factor depends on the levels of the other factors}", while a \textit{factor} is defined as "\textit{A controllable experimental variable that is thought to influence the response}". According to the so-called \textbf{sparsity-of-effects principle}, the low-order interactions monumentally dominate the high-order ones. This concept is implemented in the PCE equations by manipulating the multi-index set in Eq.(\ref{eq:multi_i}). Altering the multi-index set clearly changes Eq.\,(\ref{eq:main_pc}) and the subsequent ones. The new hyperbolically-truncated multi-index set is referred to as $\mathcal{I}_{m,p,q}$. The new term in $\mathcal{I}$ defines the so-called '$q$-norm', a quantity explained in\,\ref{app:3}. It defines a quasi-norm in the 'probability space', as the probability space is also a metric space. The '$q$-norm' is selected between 0 and 1, with 1 being not truncated.

Next, a method used in machine learning, the \textbf{least-angle regression} method (LAR), is employed in the context of PCE. LAR employs a routine of iterating over the remaining chaos basis functions to select the set that has the highest influence on the model response. The basis will be \textit{filtered} according to their contribution to the system response regardless of their order of interaction. The LAR algorithm was originally proposed by Efron\,\cite{efron2004}, while\,\cite{hastie2009elements} is a descent of the least-square regression method used to solve LLSE. The LAR algorithm is not used to compute the chaos expansion coefficients but to select the basis functions. Then, the LSE is used to compute the coefficients. The resulting LAR truncated multi-index set is referred as $\mathcal{I}_{m,p,q}^*$. Executing the LAR algorithm results in a number of solutions (many possible basis functions sets). The selected LAR model chosen that yields the minimum leave-one-out error $L_{\text{oo}}$ is explained later. The coefficients are then calculated using the ordinary LSE regression. If the $L_{{oo}}$ does not reach the required threshold accuracy, new samples are added and the algorithm is repeated. The algorithm chart is detailed in\,\citep{yu2015advanced}.

\subsection{Cross-Validation}\label{val}

Cross-validation is a technique employed to assess the quality of the PCE meta-model. The basic idea is to decompose the (input/output) data into $K$ sets. $K-1$ sets are used to build the fit model and the error is calculated by predicting the remaining set not included in the fitting calculations. This method is called the $K$-fold cross-validation technique\,\cite{hastie2009elements}. When $K$ equals the cardinality of the design of experiment $N$, the validation method is called the leave-one-out cross-validation (LOOCV). The implemented algorithm selects a single instance $\bm{\xi^k}$ from the input set and computes the meta-model of the remaining set, \textit{i.e.}, $\mathcal{M} \left( \bm{\xi} - \{\bm{\xi^k}\}  \right) = \mathcal{Y}^k$. The algorithm iterates over each of the sets and the leave-one-out error is calculated according to
\begin{align*}
e_{{rr}_{\text{LOO}}}=\frac{1}{N}\sum_{k=1}^{N}\left( \mathcal{Y} -\mathcal{Y}^k \right)^2
\end{align*} 
\noindent A reliable PCE meta-model requires a leave-one-out error $e_{{rr}_{\text{LOO}}}$ in the order of $10^{-2}$\,\cite{sudret2000stochastic}. 

\section{Results} \label{sec5}
The 10-dimensional input data were generated using a quasi-random scheme, called the \textbf{nested} latin hypercube sampling (NLHS).
Firstly, latin hypercube sampling (LHS) is an efficient sampling scheme proposed by McKay\,\citep{doi:10.1080/00401706.1979.10489755} that offers better representation of the input data compared to MC-based sampling methods\,\citep{blatmanPhD}. It offers similar performance compared to MC-based schemes in the sense that the latter requires a larger number of samples and consequently a larger number of simulations to provide the same quality as the LHS. The nested LHS (NLHS) allows the initial size of the design of experiment to be increased in order to fulfill the $e_{{rr}_{\text{LOO}}}$. One does not know in advance how many simulations must be performed in order to build a reliable statistics. It is probable (as in this paper) that the number of simulations must be increased to meet the cross-validation conditions. The NLHS is one validated method\,\cite{yu2015advanced} to do this.  

\begin{figure*}[bt]
\centering
\subfigure[Estimated coefficients (vertical cuts) for each meta-model of the electric field homogeneity $\mathcal{Y}_E$, created by the least-angle regression algorithm. On the horizontal axis, the iteration number is shown. The optimum coefficients set is shown by the vertical cut of the red-dashed vertical line with respect to the leave-one-out cross validation.]
{\includegraphics[width=0.45\textwidth]{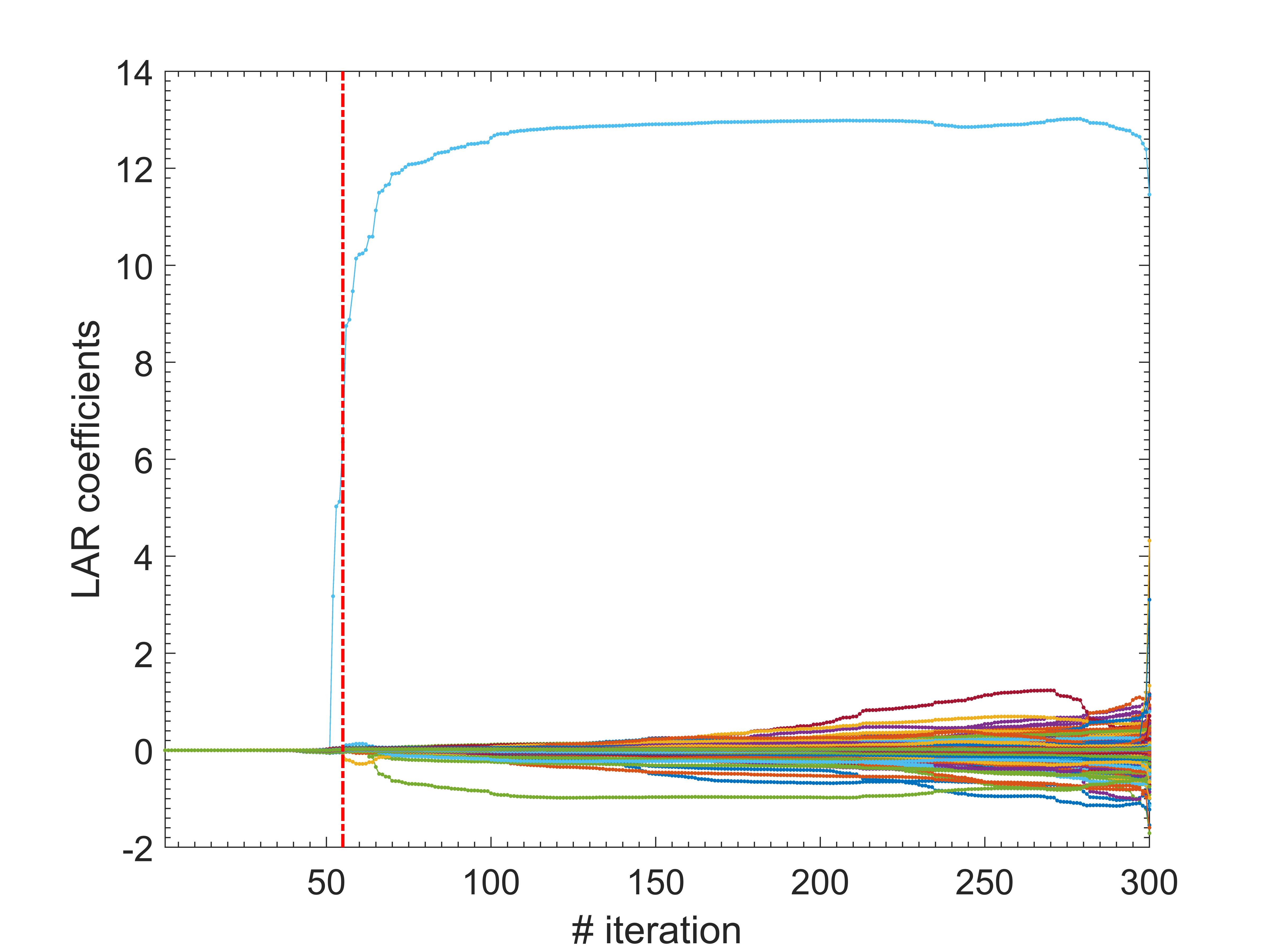}}
\hspace{0.5cm}
\subfigure[Estimated coefficients (vertical cuts) for each meta-model of the magnetic field homogeneity $\mathcal{Y}_H$, created by the least-angle regression algorithm. On the $x$-axis, the iteration number is shown. The optimum coefficients set is shown by the vertical cut of the red-dashed vertical line with respect to the leave-one-out cross validation.]  
{\includegraphics[width=0.45\textwidth]{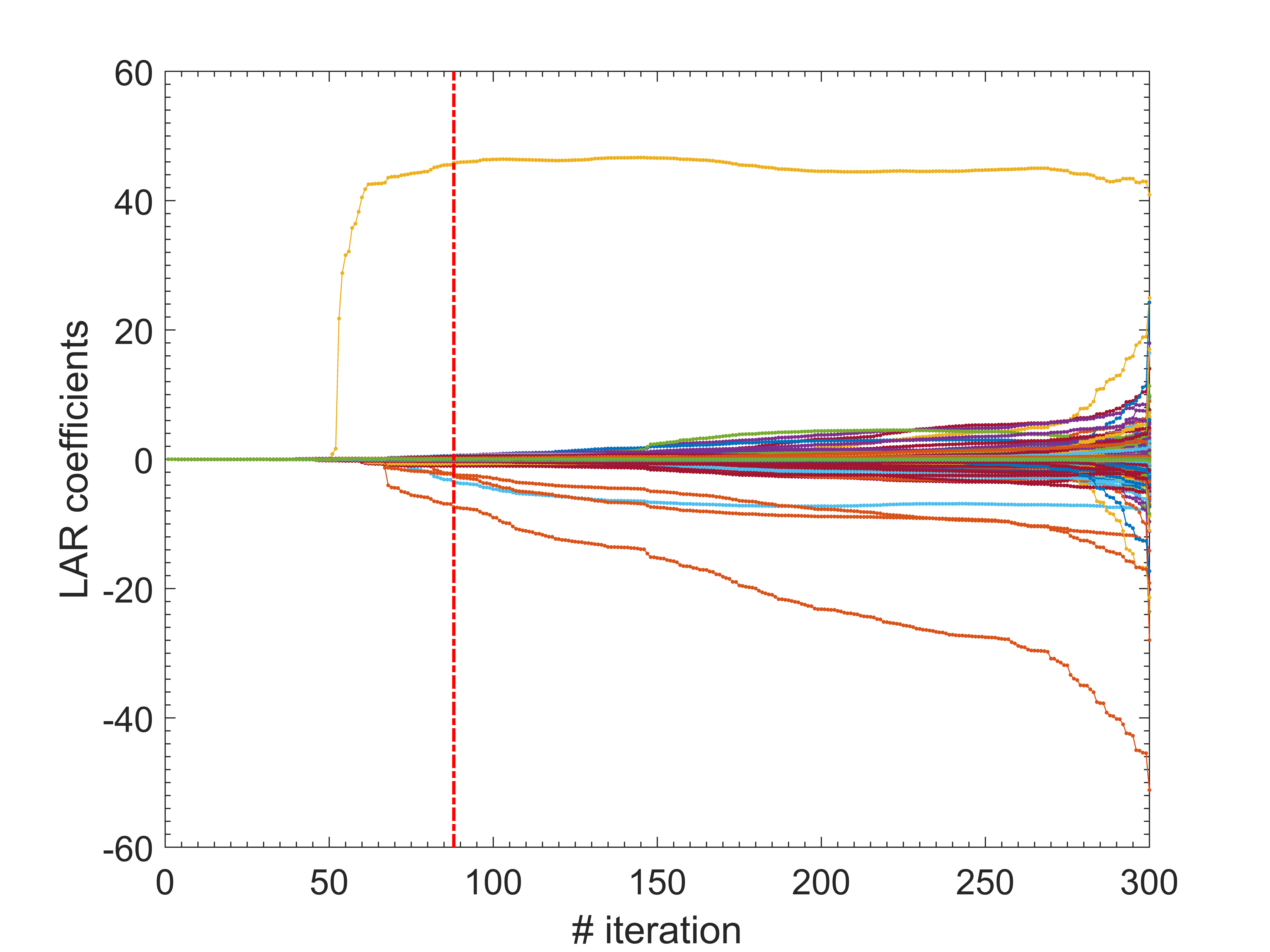}}
\caption{\label{fig:lar} The sparsity of the PCE depends on the least-angle regression algorithm which produces multiple meta-models for the electric and magnetic field homogeneities. The selected meta-model will be used to rebuild the multi-index set and then select the basis functions.}
\end{figure*}

At first, 100 full wave simulations have been carried out, then the number of simulations was increased to 300, until the leave-one-out error threshold has been reached. A $q$-norm $q$ of 0.4 has been selected, as suggested in\,\cite{blatmanPhD}. The PCE expansion was truncated at the $p=11^\text{th}$-order. Applying the LAR algorithm results in 300 meta-models with the coefficients for the electric and magnetic fields,  respectively, shown in Fig.\,\ref{fig:lar}. On each set of LAR coefficient the leave-one-out cross-validation is performed and the meta-model with the lowest value is select to be the optimum solution as shown by the dashed red vertical lines in Fig.\,\ref{fig:lar}. For the electric field, only 55 basis functions are used to produce a final $e_{{rr}_{\text{LOO}}} = 0.0264$, while for the magnetic field 84 basis functions were computed to provide a $e_{{rr}_{\text{LOO}}} = 0.0452$. This means that in total 168 full-wave simulations would be sufficient to produce a Monte-Carlo equivalent result. With the optimum meta-model in hand, the new, LAR-based, multi-index set $\mathcal{I}_{m,p,q}^*$ is calculated and consequently the basis functions $\bm{\Psi}$ are constructed. The expansion coefficients can be easily computed by applying Eq.\,(\ref{eq:olse}).

\begin{figure*}[bt]
\centering
\subfigure[Estimated statistical distribution of the electric field homogeneity $\hat{\mathcal{Y}}_E= f_{E_\perp}^{\text{int}}$ calculated using the sparse generalized PCE method with an expansion order 12 fitted to a Gaussian distribution. $10^4$ samples of the PC basis have been used for the construction of the distribution. The result of fitting the low-order MC $\mathcal{Y}_E$ with a Gaussian distribution is displayed as well.]
{\includegraphics[width=0.45\textwidth]{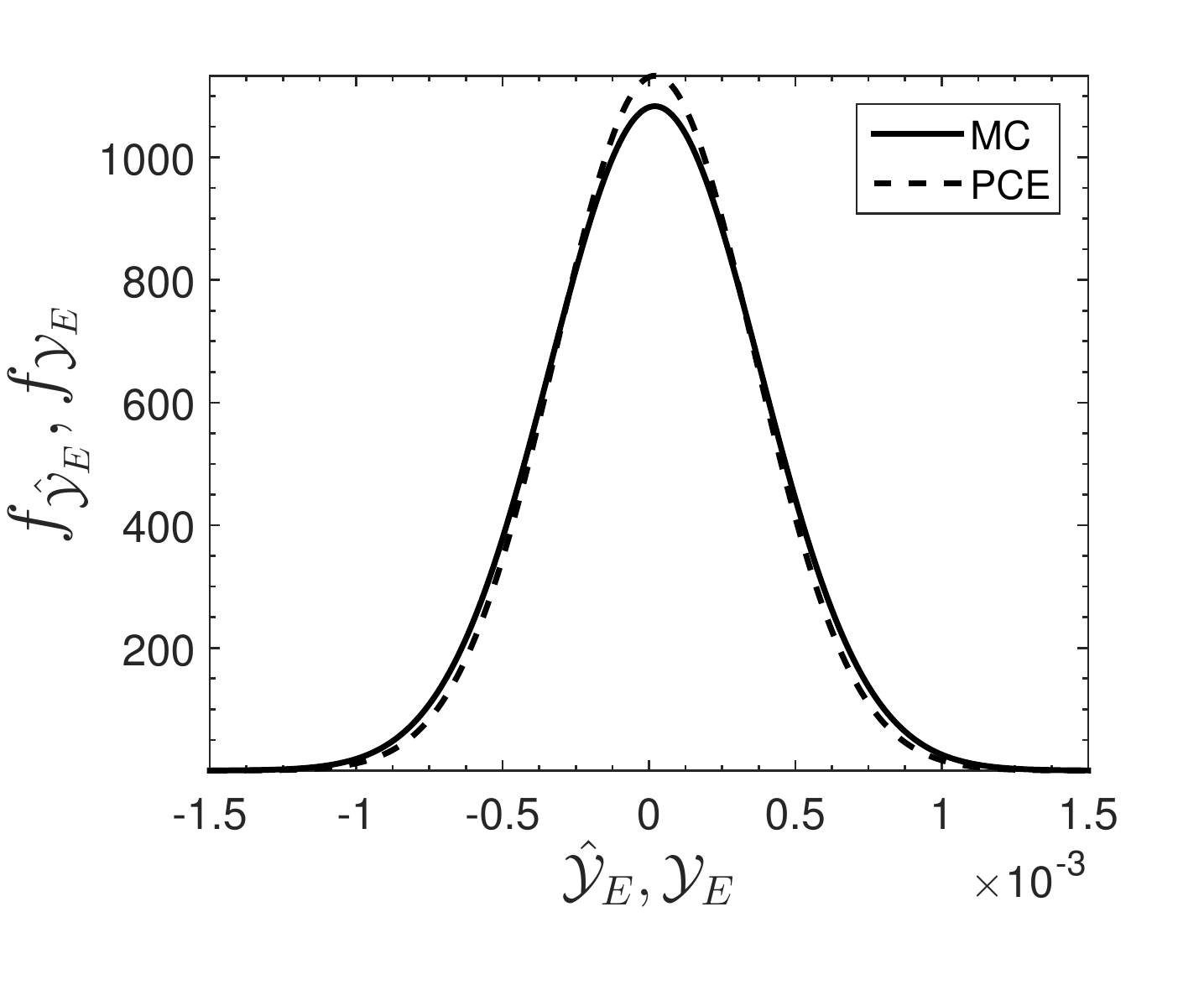}}
\hspace{0.5cm}
\subfigure[Estimated statistical distribution of the magnetic field homogeneity $\hat{\mathcal{Y}}_H= f_{H_\perp}^{\text{int}}$ calculated using the sparse generalized PCE method with an expansion order 12 fitted to a Gaussian distribution. $10^4$ samples of the PC basis have been used for the construction of the distribution. The result of fitting the low-order MC distribution $\mathcal{Y}_H$ with a Gaussian distribution is also shown.]  
{\includegraphics[width=0.45\textwidth]{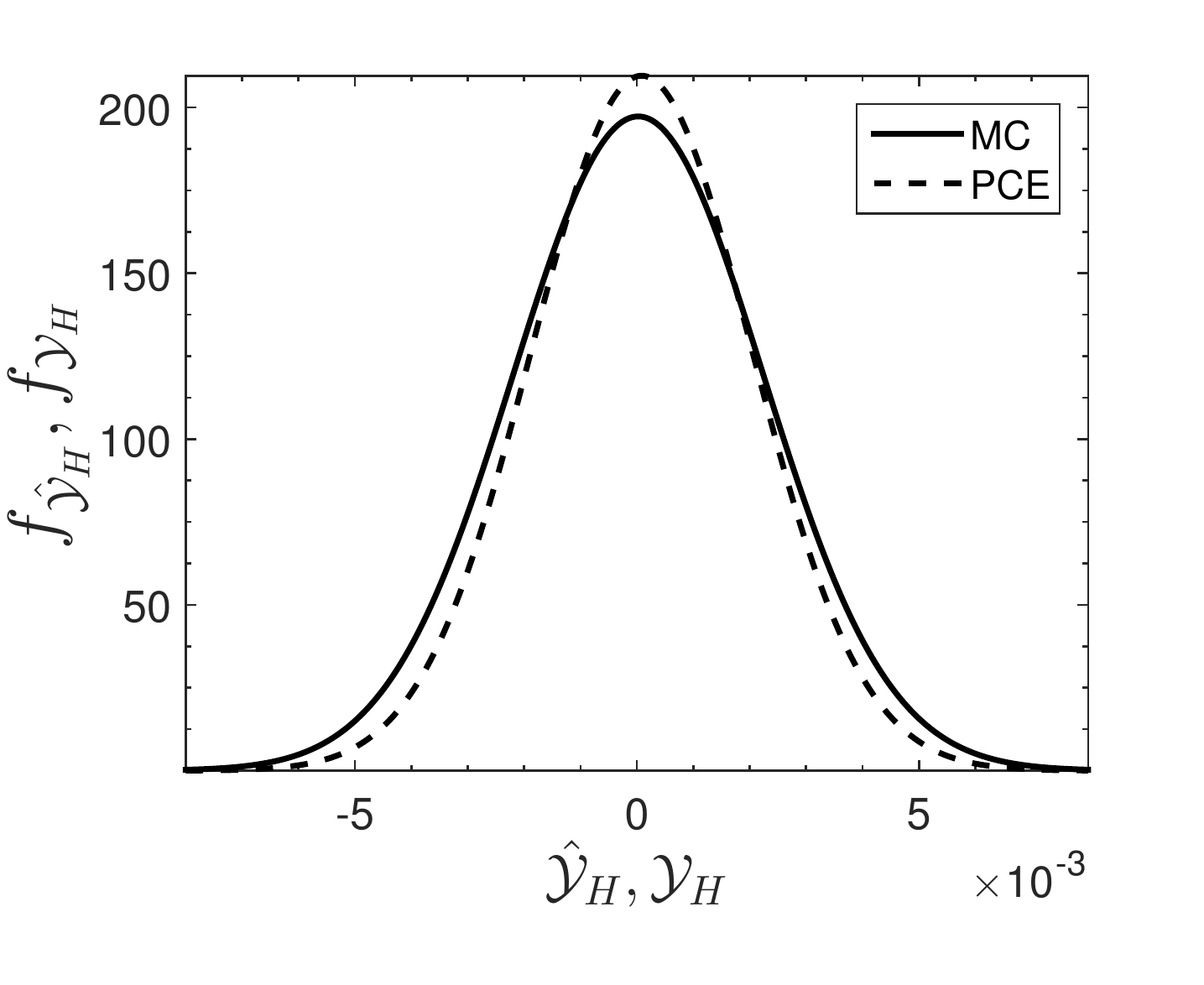}}
\caption{\label{fig:pdf} Comparison of the statistical distribution of the estimated homogeneities of the electric and magnetic field computed using the sparce PCE and the low-order MC method. The quantitative results are summarized in Table\,\ref{tab:statistics}.}
\end{figure*}

The probabilistic performance of the RF Wien filter is shown in Fig.\,\ref{fig:pdf}. These are not the probability distributions; the results were fitted with a Gaussian distribution with arbitrary mean values and standard deviations, and the parameters are summarized in Table\,\ref{tab:statistics}. The magnetic field undergoes much faster and stronger variations than the electric field. This is clearly obvious from Fig.\,\ref{fig:eh_perp}. As a result, the values of the homogeneity integrals for the magnetic field will me more difficult to detect and to estimate. The number of basis functions to span the variation of the magnetic field was higher. Even in Fig.\,\ref{fig:pdf}, the electric field fits better to the 1000 simulations than the magnetic field. 
\begin{table}[tb]
\renewcommand{\arraystretch}{1.3}
\centering
\caption{Mean and standard deviations of the PCE and MC-based system response for $f_{E_\perp}^{\text{int}}$ and $f_{H_\perp}^{\text{int}}$.}
\begin{tabular}{lcc}\hline \hline
					& Mean $\mu$ 				& Width $\sigma$\\\hline
${\mathcal{Y}_E}^{\text{(MC)}}$		& $(2.03 \pm 6.81) \times 10^{-5}$ 	& $(4.95 \pm 0.96) \times 10^{-4}$\\
${\hat{\mathcal{Y}}_E}^{\text{(PCE)}}$	& $(2.39 \pm 6.60) \times 10^{-5}$ 	& $(4.87 \pm 0.93) \times 10^{-4}$\\\\
${\mathcal{Y}_H}^{\text{(MC)}}$		& $(1.54 \pm 0.23) \times 10^{-4}$ 	& $(3.22 \pm 0.04) \times 10^{-3}$\\
$\hat{\mathcal{Y}}_H^{\text{(PCE)}}$	& $(1.91 \pm 0.38) \times 10^{-4}$ 	& $(3.08 \pm 0.06) \times 10^{-3}$\\\hline \hline
\end{tabular}
\label{tab:statistics}
\end{table}

According to the results of the PCE simulations (see Table\,\ref{tab:statistics}), the electric field maintains a high field homogeneity compared to the ideal (no-uncertainty) model. However, the magnetic field seems to be more sensitive to mechanical variations by about a factor of 10 than the electric field. 

\section{PC-based sensitivity analysis} \label{sec6}
To conclude this analysis, a sensitivity analysis is indispensable, which will allow us to identify the most influential parameters on the performance of the device. The results may be used by the mechanical engineers during the assembling of the RF Wien filter.  

The basic idea is to decompose the variance of the output ($\mathcal{Y}_E$ and $\mathcal{Y}_H$) as a function of the contribution of each variable and possibly their combination. This is called the ANalysis Of VAriance, or ANOVA. The independence of the random input variables and the orthonormality of the PCE, permits the direct computation of one known sensitivity scheme,  called the Sobol sensitivity via the Sobol decomposition at zero-cost\,\cite{Deman2016156}. Zero-cost means that no additional computation is required. The PC coefficients can be used directly.

If $\mathcal{Y}$ is the \textit{total} output, then $\mathcal{Y}_{\xi_i}$ denotes the output produced by the random variable $\xi_i$. Rewriting Eq.\,(\ref{eq:main_pc}) as a function of terms of the variable $\xi_i$ gives
\begin{equation}
\mathcal{Y}_{\xi_i}=\mathcal{M}\left( \xi_i \right) = \sum_{\bm{i} \in \mathcal{I}_{\xi_i}} \alpha_{\bm{i}}\Psi_{\bm{i}} \left(\xi \right)\,.
\label{eq:sobol_pc}
\end{equation}
Here $\mathcal{I}_{\xi_i}$ is the multi-index set of the variable $\xi_i$, \textit{i.e.}, the $i^{\text{th}}$ term in the multi-index row is non-zero. Therefore, to characterize the output related to the $\xi_i$ or any of its combinations, all that is required is the PC coefficients. If $\hat{D}$ denotes the variance of the estimated output $\hat{\mathcal{Y}}$, the partial variances corresponding to the random input variables are denoted as $\hat{D}_{\xi_1, \dots, \xi_{10}}$. In this case, the total and partial variances are calculated, respectively, as
\begin{equation}
\begin{split}
\hat{D} = \sum_{\bm{i} \in {\mathcal{I}- \{0\}}} \alpha_{\bm{i}}^2 \,, \text{ and } \\
\hat{D}_{\xi_i} = \sum_{\bm{i} \in {\mathcal{I}_{\xi_i}}} \alpha_{\bm{i}}^2 \,.
\end{split}
\end{equation}

\begin{figure}[hbt]
\centering
\includegraphics[width=9cm]{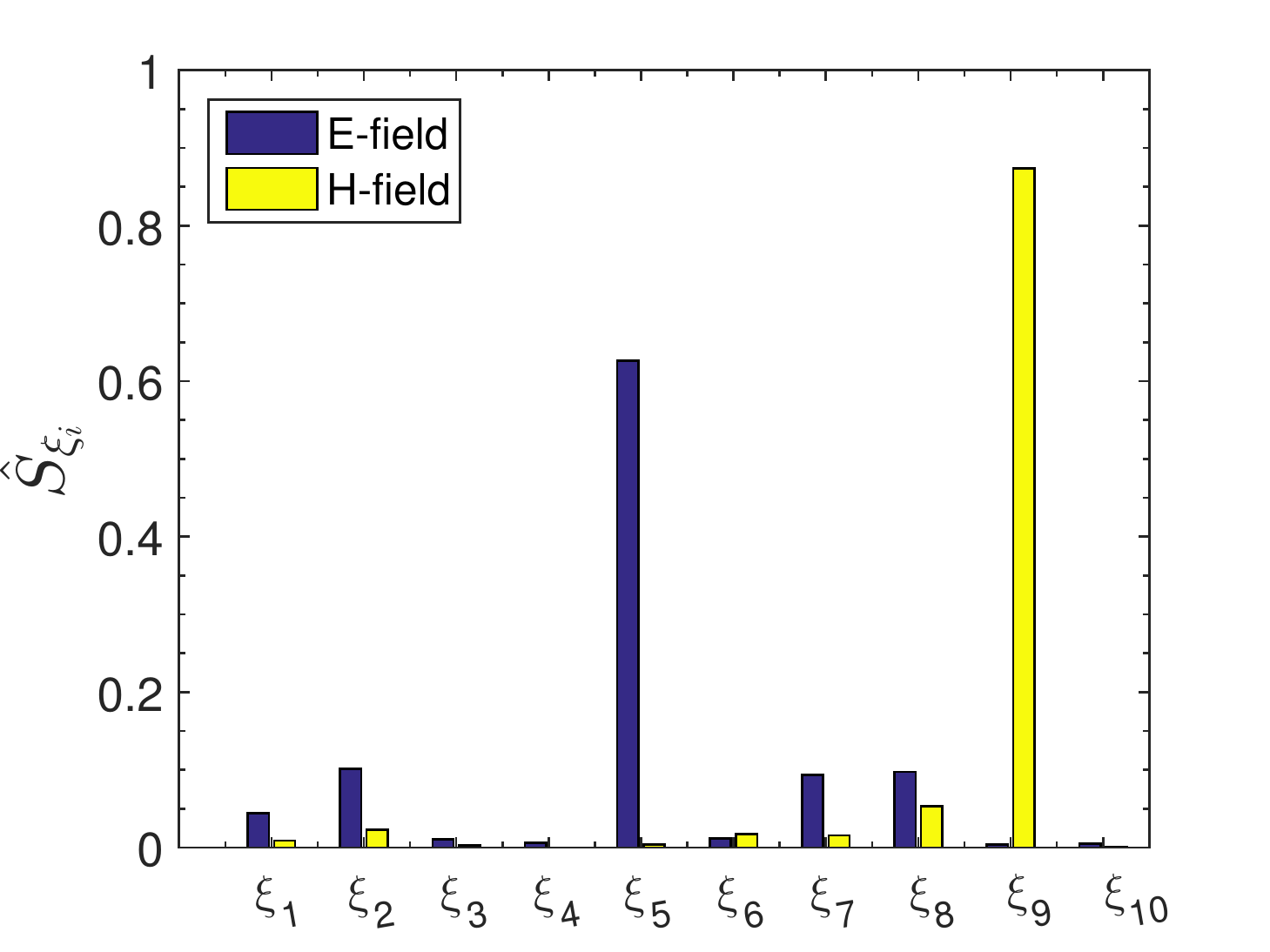}
\caption{Sobol sensitivity of the electric and magnetic field homogeneity to the design parameters of the waveguide RF Wien filter.}\label{fig:sobol}
\end{figure}

Finally, the Sobol indices corresponding to the parameters $\xi_i$ are calculated according to 
\begin{equation}
\label{eq:sens}
\hat{S}_{\xi_i} = \sum_{\bm{i} \in {\mathcal{I}_{\xi_i}}} \frac{\alpha_{\bm{i}}^2}{\hat{D}};\quad \Big\{\mathcal{I}_{\xi_i} = {\bm{i} \in \mathcal{I}: i > 0, i \ne j} = 0\Big\}\,.
\end{equation}
Only the first order Sobol indices are shown; they constitute 97 \% of the total contribution over the higher order indices, as can be seen in Fig.\,\ref{fig:sobol}. The results of evaluating Eq.\,(\ref{eq:sens}) for the electric and magnetic fields  are shown, respectively, by the blue and yellow bars in Fig.\,\ref{fig:sobol}. On the $x$-axis, the uncertain parameters, \textit{i.e.},  $\xi_1,\ldots, \xi_{10}$ are shown. Within the $\pm \SI{0.1}{mm}$ manufacturing uncertainty and according to the PCE simulations, the most influential variables on the electric field  homogeneity are $\xi_1$, $\xi_2$, $\xi_5$, $\xi_7$, and $\xi_8$. The tolerances in the electrodes lengths, $\xi_1$ and $\xi_2$, are responsible for 5 to 10\% deviation of the field homogeneity from the mean value. Because $\hat{S}_{\xi_1}$ and $\hat{S}_{\xi_2}$ have different levels, this implies that the electrode displacement contributes to the performance. The widths of the plates does not have a noticeable influence on the homogeneity of the electric field. The rotation of the upper electrode with respect to the $xy$-plane has the highest sensitivity value (62\%). $\xi_6$ also represents the rotation of the second electrode with respect to the $xy$-plane but with minimum effect. Electrode misalignments and the rotation of the ferrite structure in the $xz$-plane contribute by up to 20\%. 

The ferrites, due to their high magnetic permeability $\mu$ value, flatten the magnetic field lines inside the RF Wien filter. At the corresponding frequencies, the ferrites have a larger impact on the magnetic field than on the electric field. The Sobol index is expected to be high for the values related to the ferrites. The sensitivity analysis for the magnetic field shows that $\xi_9$ is the most important one, reaching a value of 85\%.  The parameter $\xi_9$ represents the rotational misalignment of the ferrite structure in the $xy$-plane. If the angular alignment of the ferrite structure could be improved by about a factor of 3 from $\SI{1}{mrad}$ to $\SI{0.3}{mrad}$, a substantial improvement of the standard deviation of $H_\perp$ is to be expected. The rotation of the electrodes in the $xy$-plane, as well as in the $xz$-plane affect slightly the standard deviation of $H_\perp$ by about 10\%. $H_x$ is the main source of parasitic magnetic field, though it is very small compared to the main field, it does not cancel out when integrating along the longitudinal axis, but rather accumulates. 

Similar variables, \textit{e.g.}, $\xi_5$ and $\xi_6$, that represent the same physical quantity, should not necessarily possess the same Sobol sensitivities (see Fig.\,\ref{fig:sobol}). 
The electromagnetic effect of a rotation of the upper or the lower electrode is exactly the same because a parallel-plate waveguide is symmetrically invariant with respect to the $xy$-plane. When modeling the uncertainties, \textit{e.g.}, $\xi_5$ and $\xi_6$ are allowed to vary independently within their error margins [and the same is true for the other pairs $(\xi_1, \xi_2)$, $(\xi_3, \xi_4)$, and $(\xi_7, \xi_8)$]. In this way, a misalignment (breaking of parallelism) of the electrodes in the \textit{xy}-plane can be modeled using only two variables. The stochastic independence of $\xi_5$ and $\xi_6$ makes sure that the electromagnetic response from the two electrodes is different, unless the electrodes are \textit{identical}. However, if $\xi_5$ and $\xi_6$ take the same values, this means that the parallelism is conserved and the variables must have the same Sobol sensitivities. Figure\,\ref{fig:sobol} emphasizes that parallelism is the major factor in preserving the electric field. 

\section{Conclusion and outlook} \label{sec7}

This paper presents an application on sparse and non-intrusive Polynomial Chaos Expansion (PCE) as a low-cost calculation method of the electromagnetic field quality of the novel RF Wien filter. The fabrication and assembly limitations of the electrodes and the ferrite structure have been modeled and were evaluated as a function of the homogeneities of the electric and magnetic field. Because of the high-dimensionality of the problem, a sparse version of PCE has been used based on the least-angle regression method. The results were compared to low-order MC simulations, and a very good agreement between the two techniques was found. The influence of the individual parameters on the performance has been quantified based on the Sobol-sensitivity analysis. 

It has been found that the parallelism of the electrodes in the \textit{xy}-plane is the most crucial parameter that influences the performance of the electric field. The alignment of the ferrite structure in the \textit{xy}-plane was the dominant factor for the magnetic field. We do not have measurement nor simulation results in order to decide in favor of a better field homogeneity of the electric or the magnetic field. After the first EDM experiments with the RF Wien filter in 2017, we will see whether the present design criteria should be altered. For example, flat electrodes provide a more homogeneous magnetic field compared to the electric field by about a factor of 11.

The PCE theory can be applied to many problems in high-precision experiments, for which an accurate specification of uncertainties is required. Examples include,  but are not limited to the performance analysis of the design of highly-accurate BPMs, and the influence of position errors of magnetic and electric dipoles and quadrupoles on the spin-precession in storage rings.     

\section*{Acknowledgment}
This work has been performed in the framework of the JEDI collaboration (J{\"u}lich Electric Dipole moment Investigations), and is supported by an ERC Advanced-Grant of the European Union (proposal number 694340).


\appendix

\section{Building the homogeneous chaos basis}
\label{app:1}
   
In closed form, the 1D Hermite polynomials $ \mathcal{\psi}_n\left( \xi \right)$ are defined as
\begin{equation}
\mathcal{\psi}_n\left( \xi \right) = \left( -1 \right)^n \exp\left( {\frac{\xi^2}{2}} \right) \frac{d^n}{d\xi^n} \left[ \exp\left(-{\frac{\xi^2}{2}}\right) \right] \,.
\end{equation}
With respect to  the inner product in $\mathcal{L}^2$-normed Hilbert spaces, Hermite polynomials $\psi_n\left(\xi\right)$ are orthogonal with respect to the Gaussian measure, \textit{i.e.},
\begin{align*}
\big \langle \mathcal{\psi}_n\left( \xi \right)\mathcal{\psi}_m\left( \xi \right) \big \rangle
&=\frac{1}{\sqrt{2\pi}} \int_{-\infty}^{\infty} \mathcal{\psi}_n\left( \xi \right)\mathcal{\psi}_m\left( \xi \right) f_{\xi} \, d\xi\\
 &= \delta_{mn} \big \langle \mathcal{\psi}_n^2\left( \xi \right) \big \rangle\,.
\end{align*} 
Here $ f_{\xi}$ denotes the normal probability density function. 

In most cases, when more than one variable is involved, the multi-dimensional polynomial basis is required. The multidimensional Hermite polynomials are constructed using the tensor product of the 1D Hermite polynomials, via
\begin{align*}
\bm{\Psi}= \prod_{\bm{i}} \psi_{\bm{i}}\,.
\end{align*}
Here $\bm{i}$ is called the \textbf{multi-index} set (also known as tuplets); it indicates the degree of the polynomial in each of the input variables. An example of a multi-index set is located on the leftmost column of Table\,\ref{tbl:2d}. For instance, $\left(1,0\right)$ implies that the first parameter is elevated to the first order and the second to the zero$^\text{th}$ level and so on.

\begin{table}
\begin{center}
\renewcommand{\arraystretch}{1.2}
\begin{tabular}{llll}
\hline\hline
$n$ & $p$ & $n^\text{th} \psi_n$ & $\bm{i}$\\\hline
0   & 0   & 1                    & (0,0)\\
1   & 1   & $\xi_1$              & (1,0)\\
2   & 1   & $\xi_2$              & (0,1)\\
3   & 2   & ${\xi_1}^2-1$          & (2,0)\\
4   & 2   & $\xi_1\xi_2$         & (1,1)\\
5   & 2   & ${\xi_2}^2-1$          & (0,2)\\
\hline\hline
\end{tabular}
\end{center}
\caption{\label{tbl:2d} 2D PCE parameters.}
\end{table}

\section{Example}\label{app:2}
\begin{figure*}[tb]
\centering
\subfigure[10 samples with PCE of order 2.]
{\includegraphics[width=0.4\textwidth]{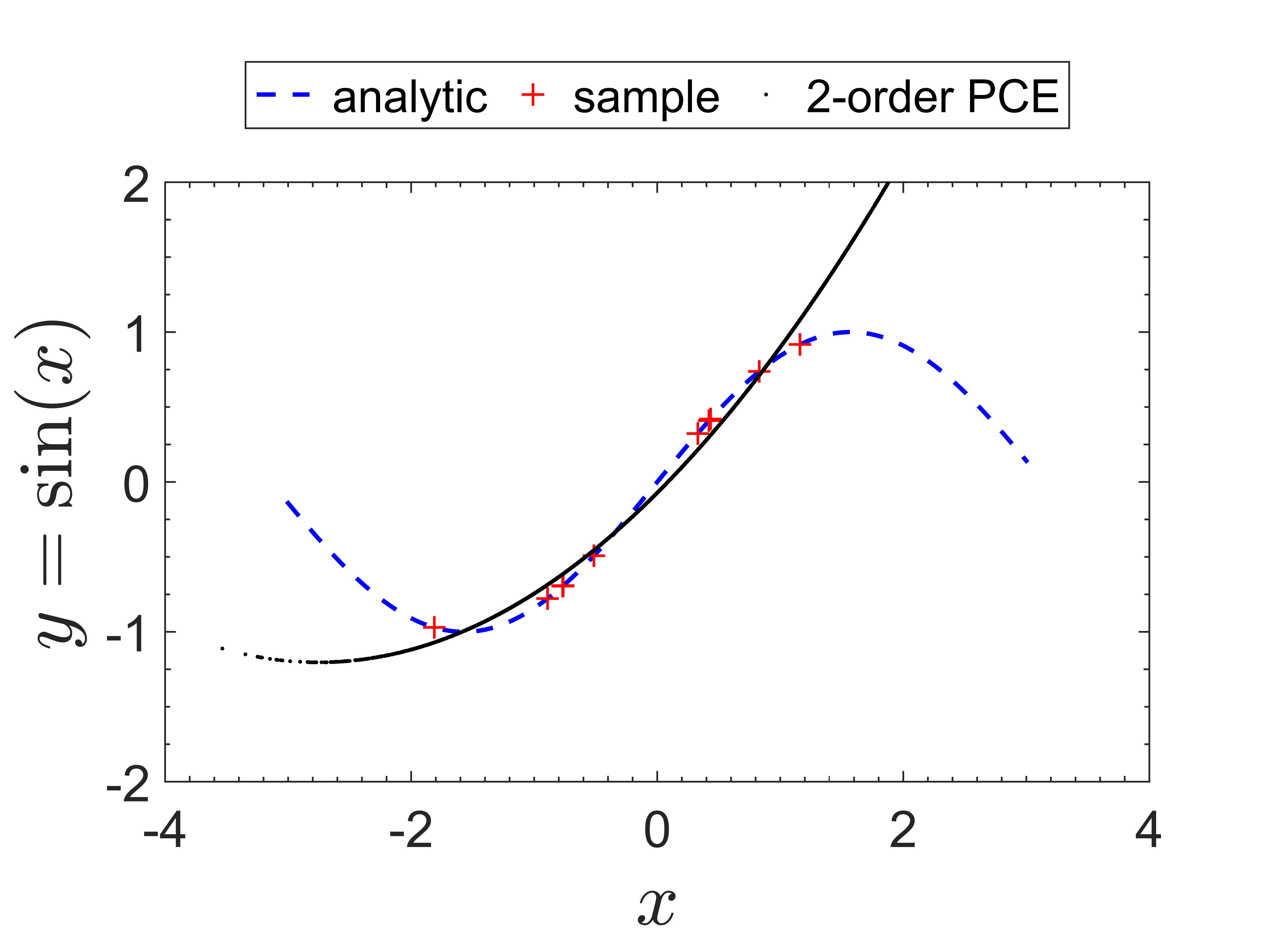}}
\subfigure[10 samples with PCE of order 4.]
{\includegraphics[width=0.4\textwidth]{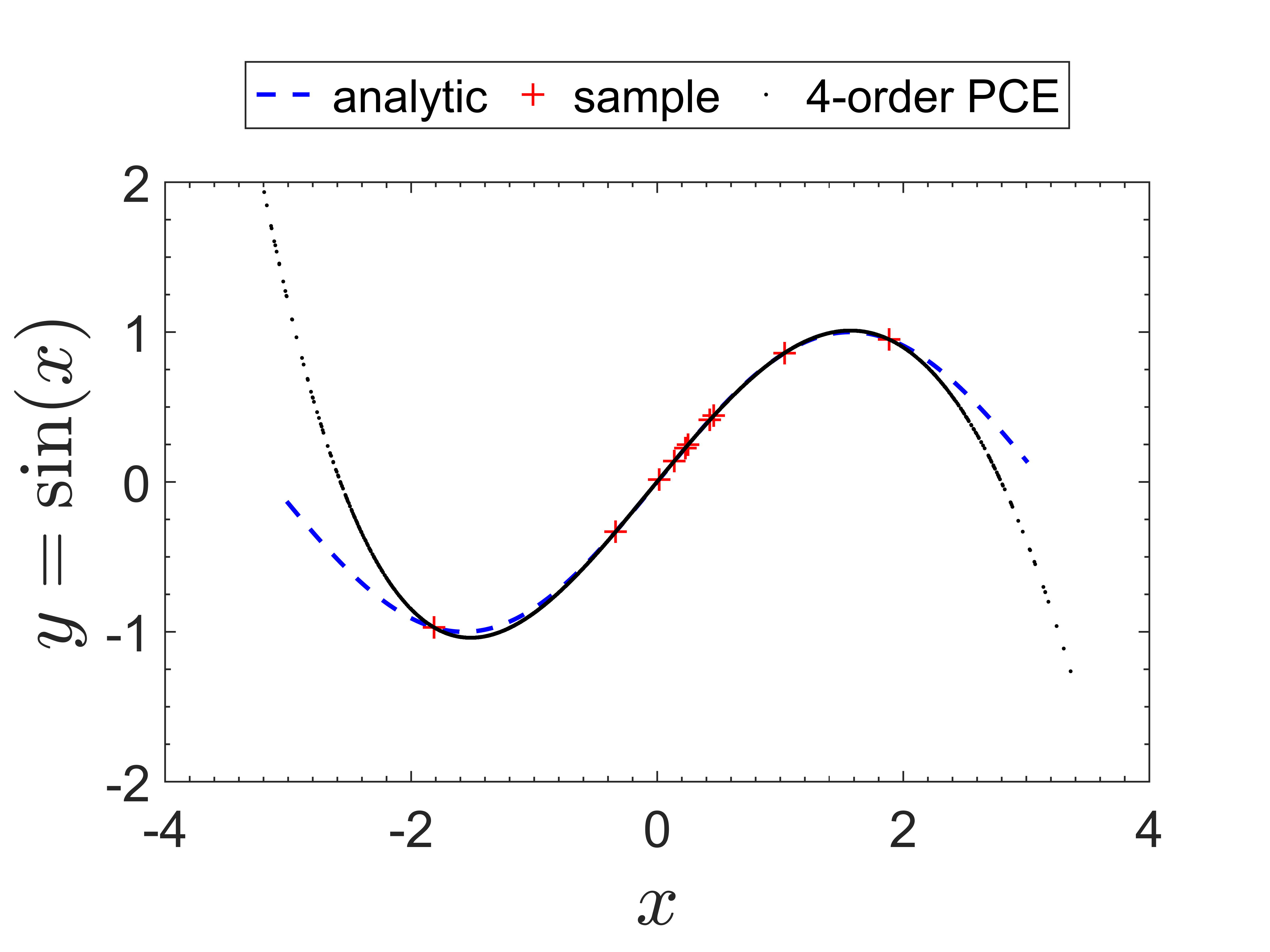}}\hspace{0.3cm}
\subfigure[10 samples with PCE of order 5.]
{\includegraphics[width=0.4\textwidth]{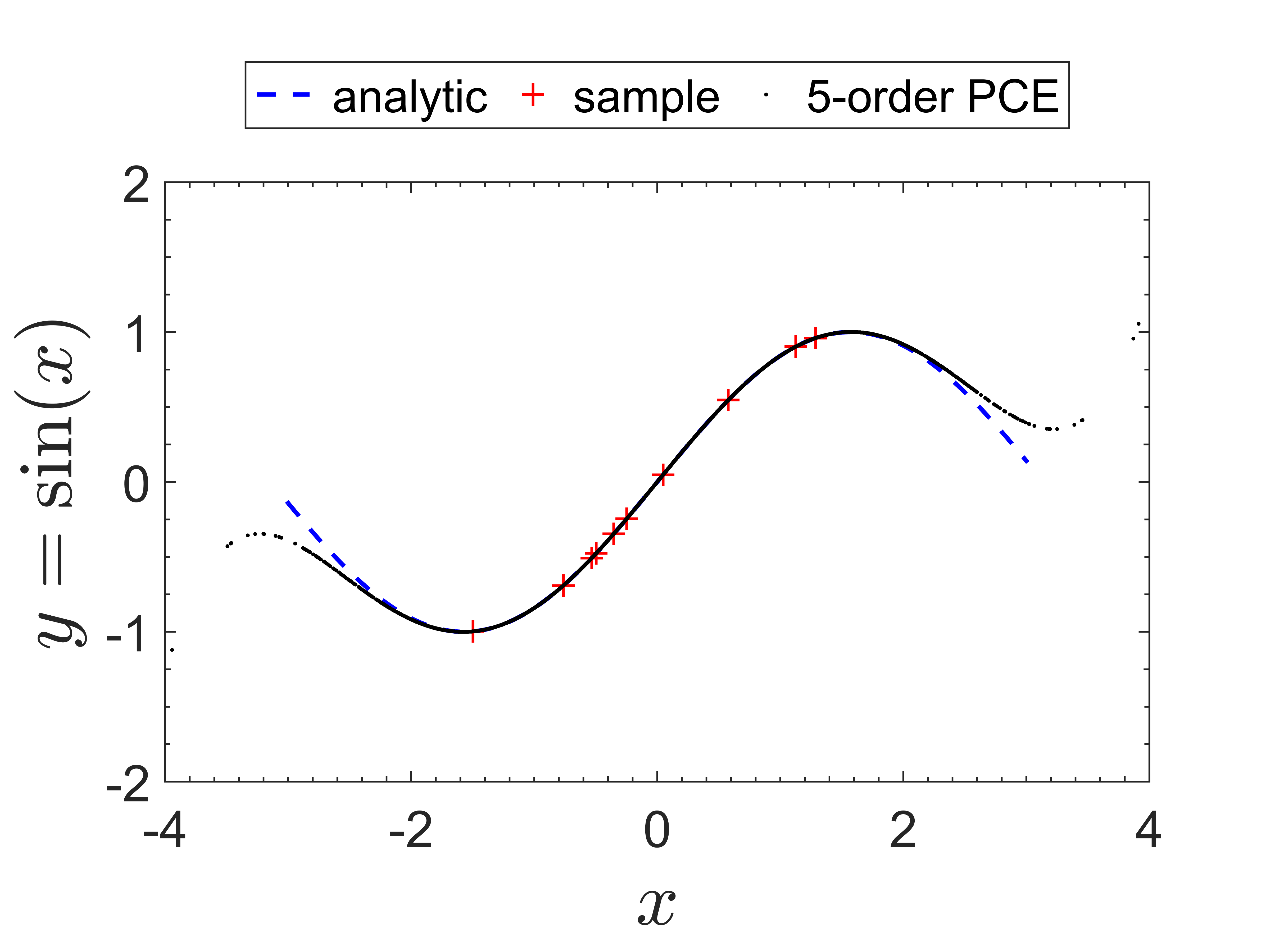}}
\subfigure[10 samples with PCE of order 8.]
{\includegraphics[width=0.4\textwidth]{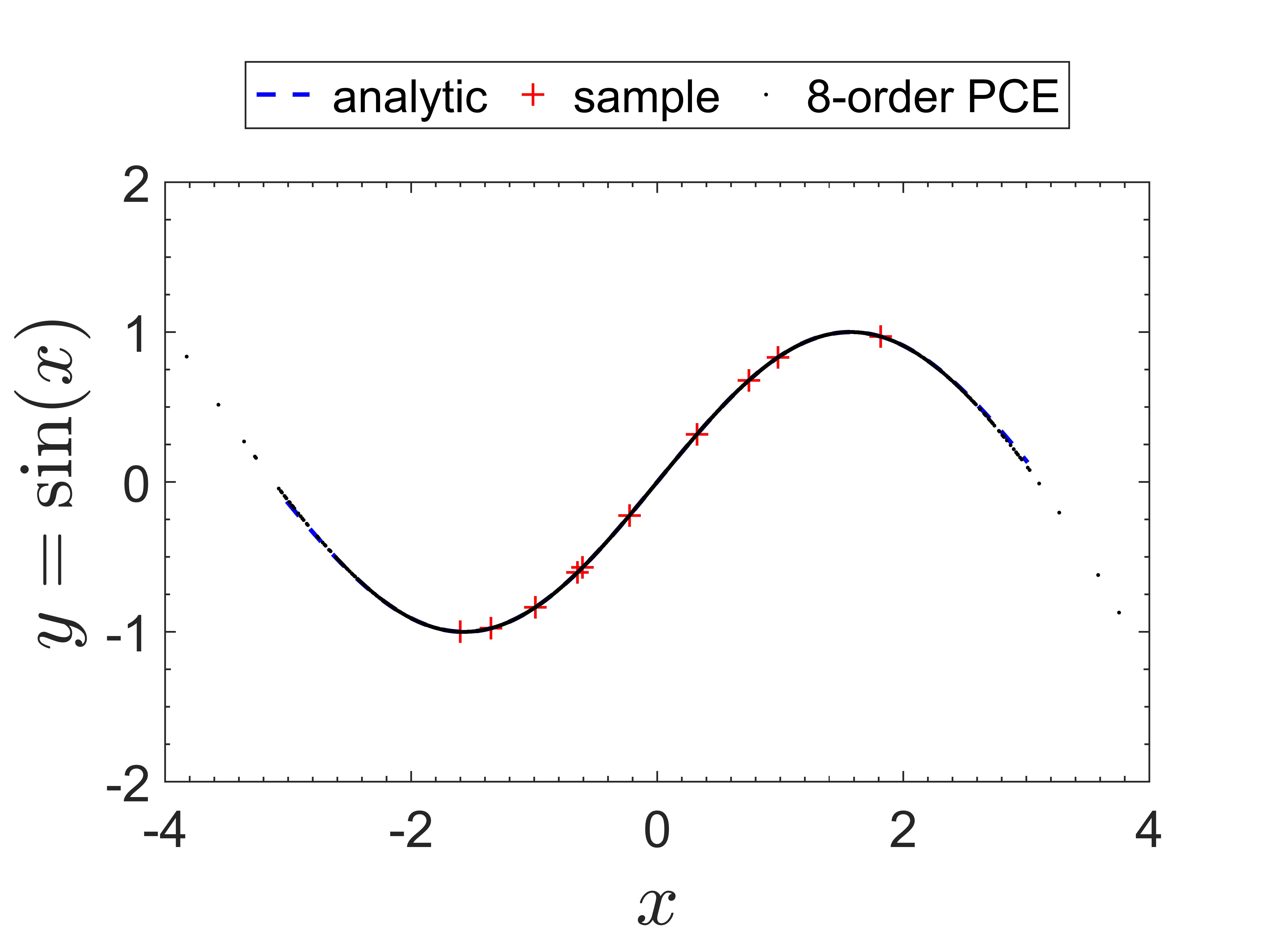}}
\caption{\label{fig:pce_ex_1}The sine function is sampled with 10 points. Increasing the PCE order, increases the fit quality.}
\end{figure*}

To further facilitate the understanding of the basic theory, in this section a simple example is provided.

Consider the function $y$, defined as
\begin{equation}
y=\sin(x)\,.
\label{sin}
\end{equation}
This function may represent the unknown dynamical behavior of an unknown system. Now, the dependent variable $x$ is not deterministic, but varies stochastically according to some distribution, say a Gaussian one. The output will then also be a stochastic variable. To build the statistics of the output, classically a Monte-Carlo simulation is conducted. Because the analytic form of the dynamics of the system is given, the performance of the PCE can be evaluated against the input analytic solution. 

The blue line in Fig.\,\ref{fig:pce_ex_1} shows the analytical solution. 10 \textit{random} sample points are created and the corresponding output is drawn (the red crosses in Fig.\,\ref{fig:pce_ex_1}). With the input and output data being present, the PCE meta-model can be constructed. As the input is a Gaussian distributed random variable, the Hermite polynomials are used to surrogate the analytical model. Figure\,\ref{fig:pce_ex_1} shows the performance of the PCE with respect to the expansion order with a constant sampling of 10 points. A $2^\text{nd}$ order expansion was clearly not sufficient to reconstruct the model response, while with the $4^\text{th}$ order expansion, the PCE could reproduce the model within the range of the sample points, but not outside this range.  On the other hand, increasing the expansion order to the $5^\text{th}$ allowed the PCE to replicate the exact model even outside the sample range but not at the edges of the function. The meta-model does have enough polynomials to predict the full response. Finally, the $8^\text{th}$ order PCE is capable to fully reconstruct the response.  
\begin{figure*}[tb]
\centering
\subfigure[The estimated probability density function (PDF) $f_y$ of both Monte-Carlo simulation and PCE.]
{\includegraphics[width=0.4\textwidth]{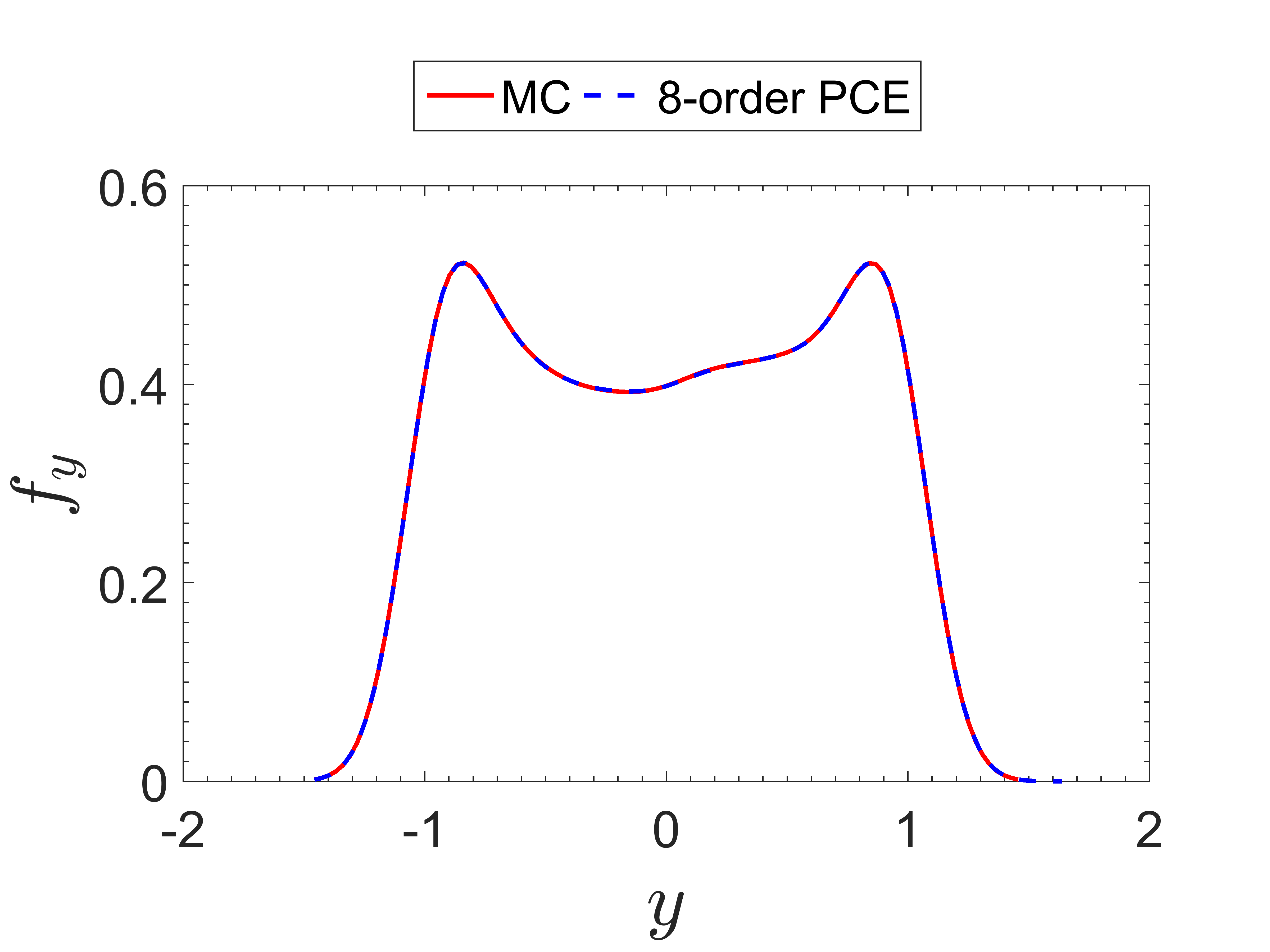}}
\hspace{0.3cm}
\subfigure[The estimated cumulative probability density function (CDF) $F_y$ of both Monte-Carlo simulation and PCE. ]
{\includegraphics[width=0.4\textwidth]{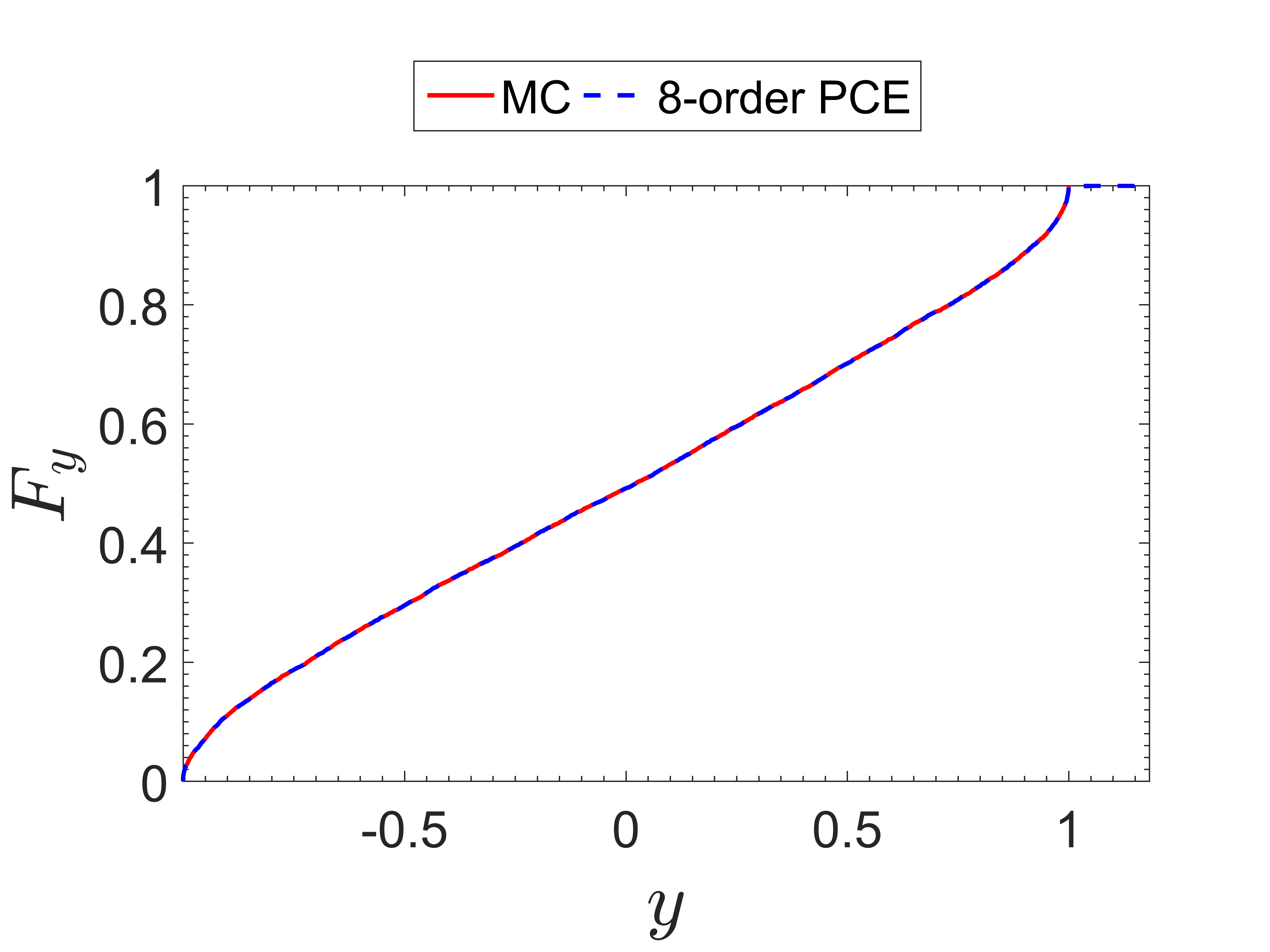}}
\caption{\label{fig:pdf_cdf} Probabilistic comparison between  MC and PCE.}
\end{figure*}

On a probabilistic basis, the statistical distributions of the output is the main criterion to consider. With a sample of 36 random points and an expansion fraction of 8, the probability density function and the cumulative density functions samples are shown in Fig.\,\ref{fig:pdf_cdf} [labels (a) and (b)], respectively. A very good congruence can be noticed proving that the PCE theory can be used to accurately surrogate models with a good accuracy.  

\section{PCE truncation}
\label{app:3}

Practically, PCE cannot extend to infinity and normally, the expansions are truncated to a degree $p$ (PC expansion order). In terms of the multi-index set $\mathcal{I}$, it is expressed in terms of dimensionality of the problem  $m$ and the expansion order $p$ as $\mathcal{I}_{m,p}$, defined as follows,  
\begin{equation}
\mathcal{I}_{m,p} = \Big\{ \bm{i} \in  \mathbb{N}^M : \| \bm{i} \|_1 \le p\Big\}\,,
\end{equation}
\noindent
where $\|\cdot\|_1$ is the 1-norm, which is simply the distance. Now, to eliminate the higher order interactions, a new form of distance is defined and applied on the multi-index set $\mathcal{I}_{m,p}$,  called the $q$-norm $\|\cdot\|_q$, which modifies the $\|\cdot\|_1$ by
\begin{equation}
\| \cdot \|_q= \left( \sum^{m} (\cdot)^q \right)^{1/q}\,.
\end{equation}

The new hyperbolic multi-index set is denoted by $\mathcal{I}_{m,p}^q$ and becomes
\begin{align*}
\mathcal{I}_{m,p}^q&= \Big\{ \bm{i} \in  \mathbb{N}^m : \| \bm{i}\|_q \le p\Big\} \\
&= \left\{ \bm{i} \in  \mathbb{N}^m : \left( \sum_{i=1}^{m} \bm{i}_i^q \right)^{1/q} \le p \right\}\,.
\end{align*}

\begin{figure}[tb]
\centering
\subfigure[$q$ = 0.8]
{\includegraphics[width=0.23\textwidth]{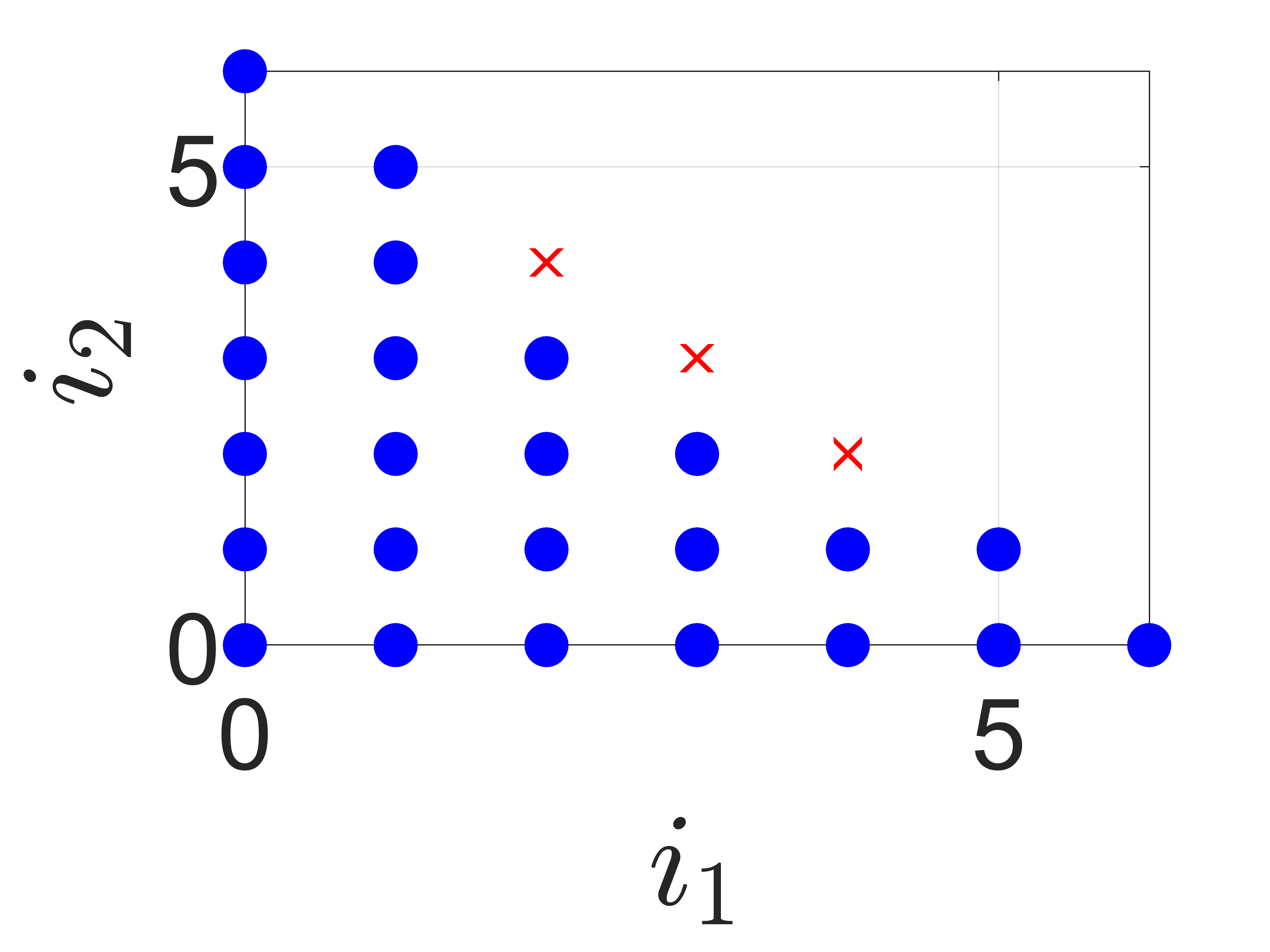}}
\subfigure[$q$ = 0.75]
{\includegraphics[width=0.23\textwidth]{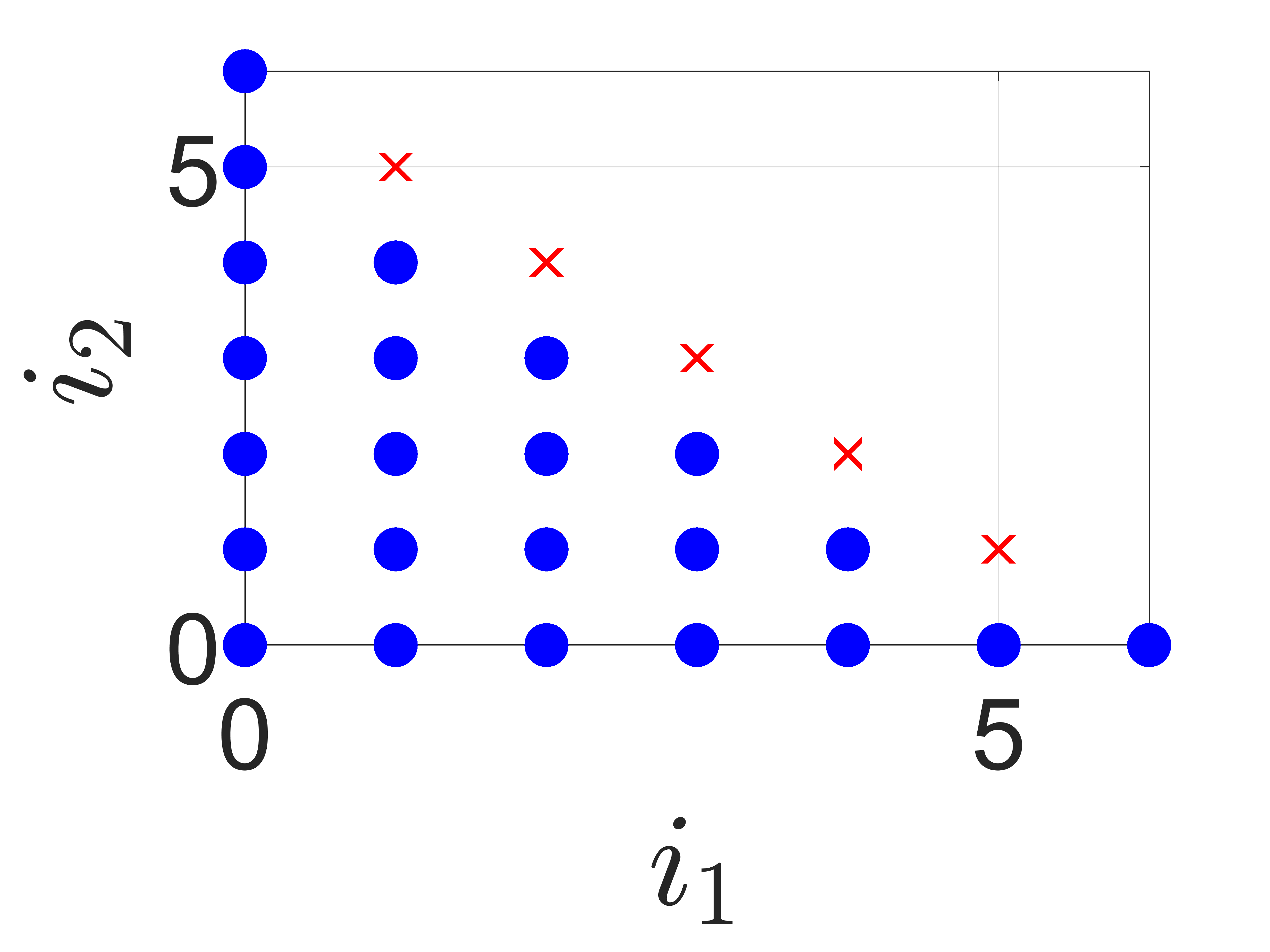}}
\subfigure[$q$ = 0.5]
{\includegraphics[width=0.23\textwidth]{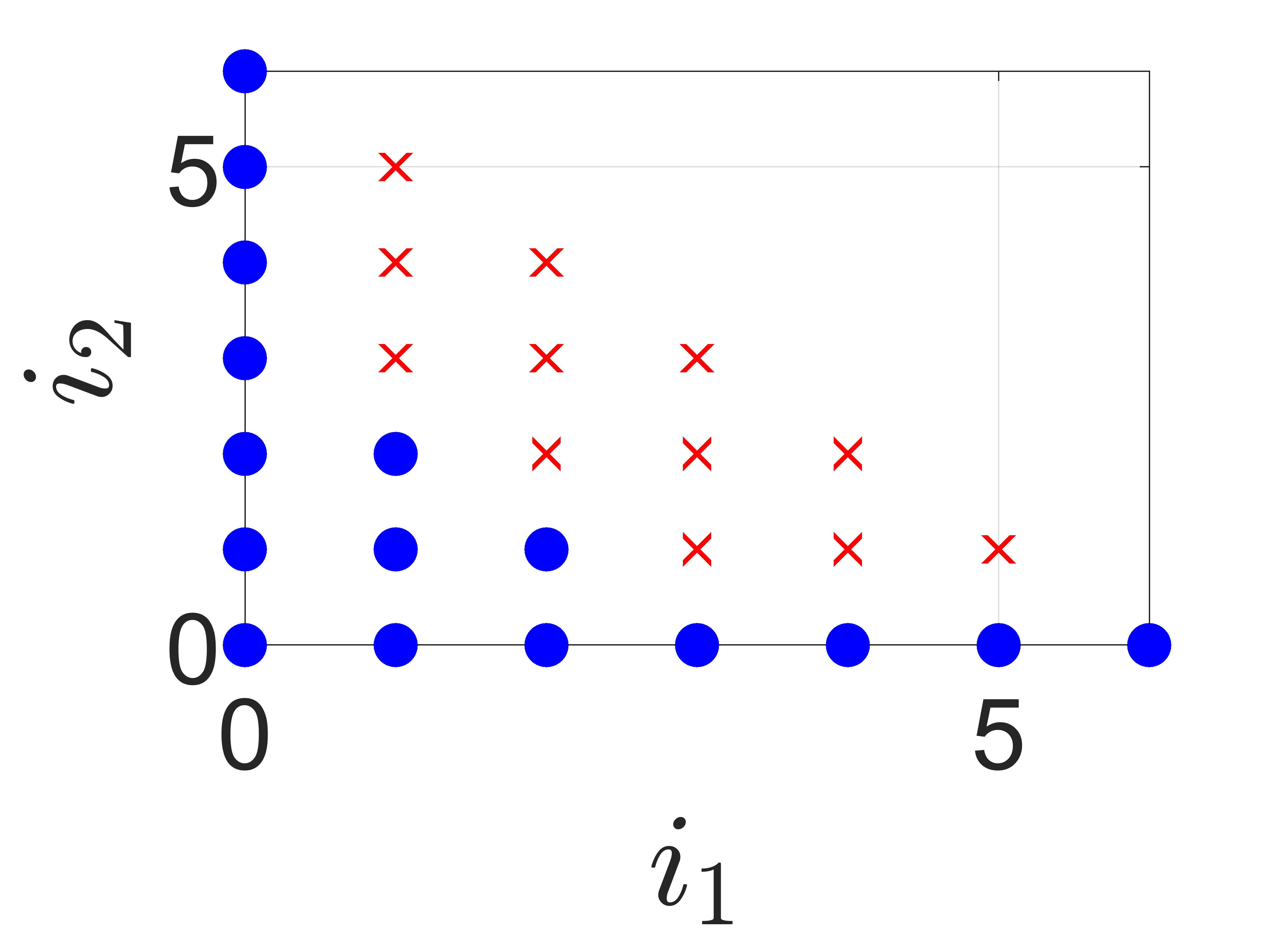}}
\subfigure[$q$ = 0.4]
{\includegraphics[width=0.23\textwidth]{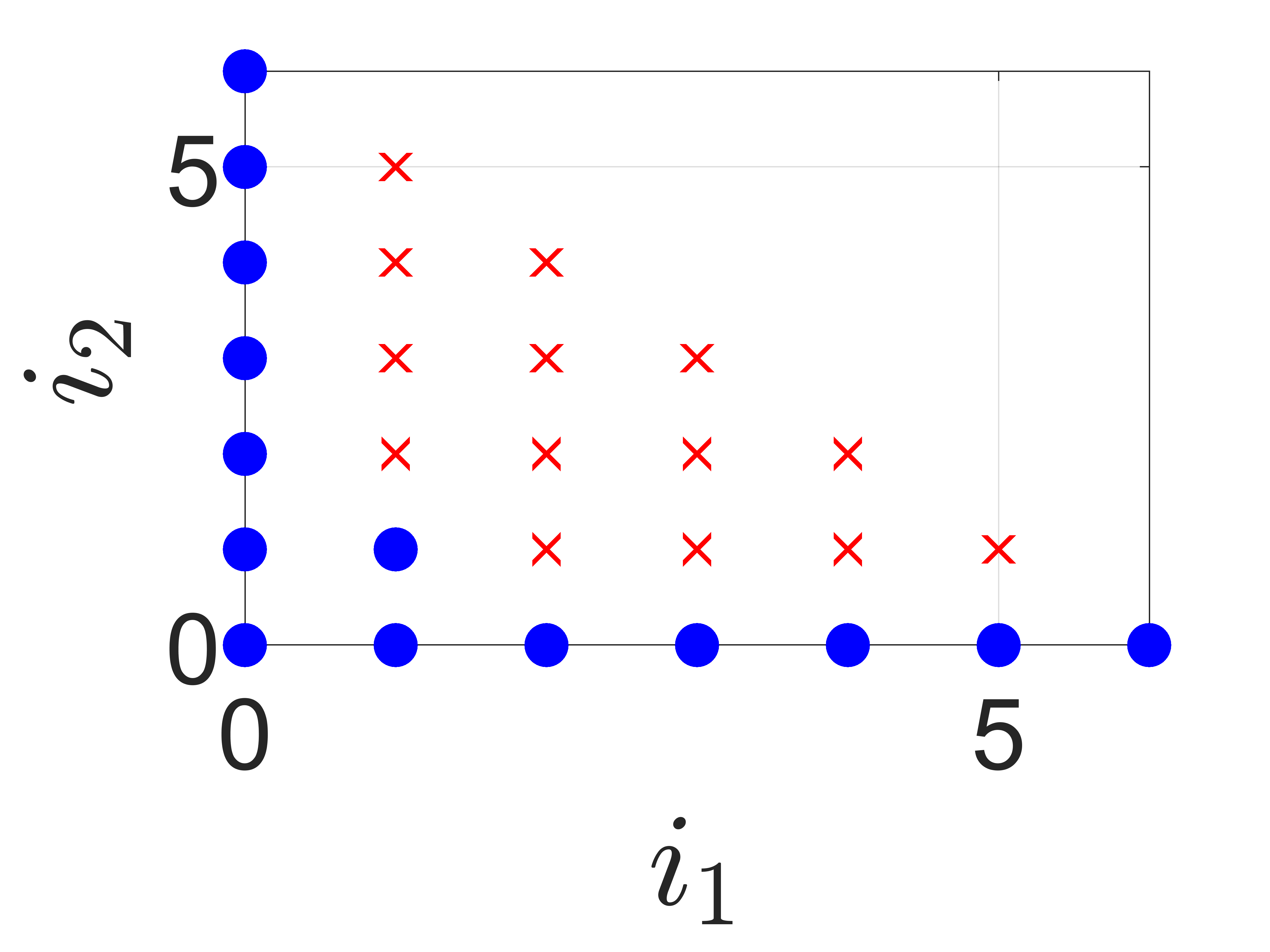}}
\caption{\label{fig:hyperbolic}Hyperbolic basis truncation  with different $q$-norms. As the $q$-norm decreases, The basis cardinality decreases. The red dots represent the full basis elements and the blue crosses indicates the selected (remaining) basis elements.}
\end{figure}

A graphical representation of the hyperbolic truncation scheme set can be found in Fig.\,\ref{fig:hyperbolic} with a $q$-norm varying from 0.8 to 0.4. The higher order interactions are taken out gradually. The polynomial terms selected for the first random variable are located on the horizontal axis, while on the vertical axis the corresponding terms of the second variables are shown. If blue dots on the outer layer are connected, the shape resembles a hyperbola, hence the name hyperbolic truncation.    

\section{The least angle regression algorithm}
\label{app:4}

The least angle regression (LAR) method is a statistical tool heavily used in machine learning and estimation theory. The LAR is a descendant of the least absolute shrinkage and selection operator method known as the (LASSO). In comparison to the full PCE, LAR-based PCE is much more difficult to implement, but in total it saves a large amount of time. With a large number of input variables $m=10$, the full PCE requires a large number of full wave simulations which is not feasible. In a similar approach conducted by Blatman et al.\,\cite{blatman2010adaptive}, the LAR-PCE based method has been implemented here with the following steps:
\begin{enumerate}
\item Set the coefficient to zero and set the residual $\bm{R}=\mathcal{Y}-\hat{\mathcal{Y}}$.
\item Find the vector (basis polynomial) $\bm{\Psi}_{\bm{i}_j}$ that is most correlated with the residual $\bm{R}$.
\item Move the corresponding coefficient $a_{\bm{i}_j}$ from 0 to $\bm{\Psi}_{\bm{i}_j}^T \bm{R}$ until another polynomial $\bm{\Psi}_{\bm{i}_k}$ has stronger correlation with the residual.
\item Move  $a_{\bm{i}_j}$ and $a_{\bm{i}_k}$ in the direction defined by their joint least square coefficient on the current residual of $\left( \bm{\Psi}_{\bm{i}j},\bm{\Psi}_{\bm{i}_k} \right)$ until some other basis has more correlation with the current residual.
\item Continue until the $P$ basis (also known as predictors) have been entered.
\end{enumerate}
\begin{figure}[tb]
\centering
\includegraphics[width=7cm]{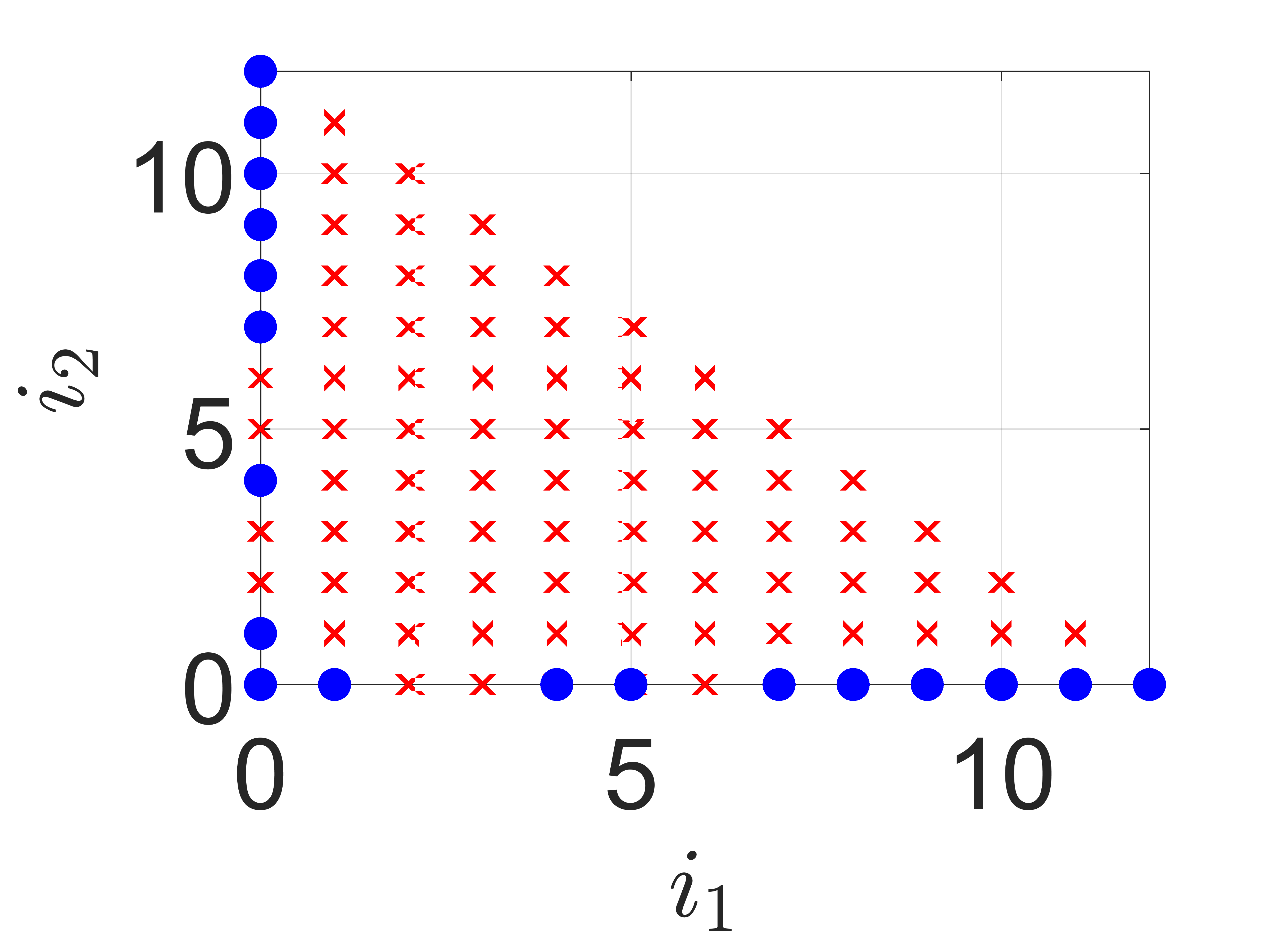}
\caption{\label{fig:h_field_lar} Applying the least-angle regression LAR to the hyperbolically truncated basis set. The order of the model is further decreased with the LAR algorithm. The red dots represent the full basis elements and the blue crosses indicate the selected (remaining) basis elements.}
\end{figure}

When applying the LAR algorithm on the magnetic field homogeneity, the selected basis functions $\mathcal{Y}_H$ are shown in Fig.\,\ref{fig:h_field_lar}. It is clear that very few polynomials have been considered in the field calculations which require much smaller number of full-wave simulations.

\bibliography{dBase}

\end{document}